\documentclass[sigconf]{acmart}

\renewcommand\footnotetextcopyrightpermission[1]{}
\setcopyright{none}
\usepackage[utf8]{inputenc}
\usepackage[T1]{fontenc}
\usepackage{amsmath}
\usepackage{amsfonts}
\usepackage{microtype}
\usepackage{graphicx}
\usepackage{algorithm}
\usepackage{algorithmic}
\usepackage{listings}
\usepackage[normalem]{ulem}
\usepackage{enumitem}
\usepackage{tabularx}
\usepackage{tcolorbox}
\usepackage{multirow}
\usepackage{wrapfig}
\usepackage{subcaption}
\usepackage{dirtree}

\tcbuselibrary{listings,breakable}
\tcbset{colback=gray!6!white, colframe=gray!40, boxrule=0.2pt, arc=1pt, left=2pt, right=2pt, top=1pt, bottom=1pt}

\DTsetlength{0.25em}{0.9em}{0.3em}{0.8pt}{3pt}

\definecolor{codegreen}{rgb}{0,0.6,0}
\definecolor{codegray}{rgb}{0.5,0.5,0.5}
\definecolor{codepurple}{rgb}{0.58,0,0.82}
\definecolor{backcolour}{rgb}{0.95,0.95,0.92}
\lstdefinestyle{mystyle}{
    commentstyle=\color{codegreen},
    keywordstyle=\color{magenta},
    numberstyle=\tiny\color{codegray},
    stringstyle=\color{codepurple},
    breakatwhitespace=false,
    breaklines=true,
    captionpos=b,
    keepspaces=true,
    numbers=none,
    numbersep=5pt,
    showspaces=false,
    showstringspaces=false,
    showtabs=false,
    tabsize=2
}
\AtBeginDocument{%
  }

\begin{document}

\title{Code2MCP: Transforming Code Repositories into MCP Services}

\author{Chaoqian Ouyang}
\authornote{Equal contribution.}
\affiliation{%
  \institution{Sun Yat-sen University}
  \country{China}}

\author{Ling Yue}
\authornotemark[1]
\affiliation{%
  \institution{Rensselaer Polytechnic Institute}
  \country{USA}}

\author{Shimin Di}
\authornote{Corresponding author; and the work was done when C. Ouyang was an intern mentored by S. Di.}
\affiliation{%
  \institution{Southeast University}
  \country{China}}
\email{shimin.di@seu.edu.cn}

\author{Libin Zheng}
\authornotemark[2] 
\affiliation{%
  \institution{Sun Yat-sen University}
  \country{China}}
\email{zhenglb6@mail.sysu.edu.cn}

\author{Linan Yue}
\affiliation{%
  \institution{Southeast University}
  \country{China}}

\author{Shaowu Pan}
\affiliation{%
  \institution{Rensselaer Polytechnic Institute}
  \country{USA}}

\author{Jian Yin}
\affiliation{%
  \institution{Sun Yat-sen University}
  \country{China}}

\author{Min-Ling Zhang}
\affiliation{%
  \institution{Southeast University}
  \country{China}}

\renewcommand{\shortauthors}{Ouyang et al.}

\begin{abstract}
The Model Context Protocol (MCP) aims to create a standard for how Large Language Models use tools. However, most current research focuses on selecting tools from an existing pool. A more fundamental, yet largely overlooked, problem is how to populate this pool by converting the vast number of existing software projects into MCP-compatible services. To bridge this gap, we introduce Code2MCP, an agent-based framework that automatically transforms a GitHub repository into a functional MCP service with minimal human intervention. Code2MCP employs a multi-agent workflow for code analysis, environment setup, tool function design, and service generation, enhanced by a self-correcting loop to ensure reliability. We demonstrate that Code2MCP successfully transforms open-source computing libraries in scientific fields such as bioinformatics, mathematics, and fluid dynamics that are not available in existing MCP servers. By providing a novel automated pathway to unlock GitHub, the world's largest code repository, for the MCP ecosystem, Code2MCP serves as a catalyst to significantly accelerate the protocol's adoption and practical application.
The code is public at \url{https://anonymous.4open.science/r/code2mcp-7E34}.
\end{abstract}




\maketitle
\pagestyle{plain}
\pagestyle{fancy}
\fancyhf{}                    
\renewcommand{\headrulewidth}{0.4pt}  
\renewcommand{\footrulewidth}{0pt}    

\section{Introduction}
\label{sec:intro}

The landscape of artificial intelligence is increasingly defined by autonomous agents that leverage Large Language Models (LLMs) to interact with external tools~\citep{wang2024survey}. To overcome the inherent limitations of LLMs in tasks requiring real-time information or precise computation, the paradigm of tool-augmented reasoning has become central~\citep{huang2024understanding, hao2023reasoning}.
Seminal works have demonstrated that models can effectively learn to invoke external functions~\citep{schickToolformerLanguageModels2023, qinToolLLMFacilitatingLarge2023, parisiTALMToolAugmented2022}.
For example, Toolformer~\citep{schickToolformerLanguageModels2023} shows that LLMs can be trained to invoke external APIs in a self-supervised manner, establishing the basic feasibility of tool use.
Later work further scales this paradigm by orchestrating larger tool and model collections, such as Gorilla~\citep{patilGorillaLargeLanguage2023} and HuggingGPT~\citep{shen2023hugginggpt}, showing that tool use can be extended beyond a small, fixed set of predefined APIs.

However, this burgeoning ecosystem faces a fundamental scalability challenge: the $N \times M$ integration problem~\citep{liangTaskMatrixAICompletingTasks2023, qinToolLLMFacilitatingLarge2023, anthropic2023mcp}.
Each of the N models or agent applications often requires a bespoke connector for each of the M tools it must access. This results in a fragmented and inefficient system where development effort is duplicated and innovation is stifled by high integration costs~\citep{quToolLearningLarge2025, shenLLMToolsSurvey2024}.
To address this, MCP is proposed as a universal standard that specifies how agents and tools should communicate, enabling an interoperable ``plug-and-play'' ecosystem~\citep{anthropic2023mcp}.
\begin{figure}[t]
  \centering
  \includegraphics[width=\linewidth]{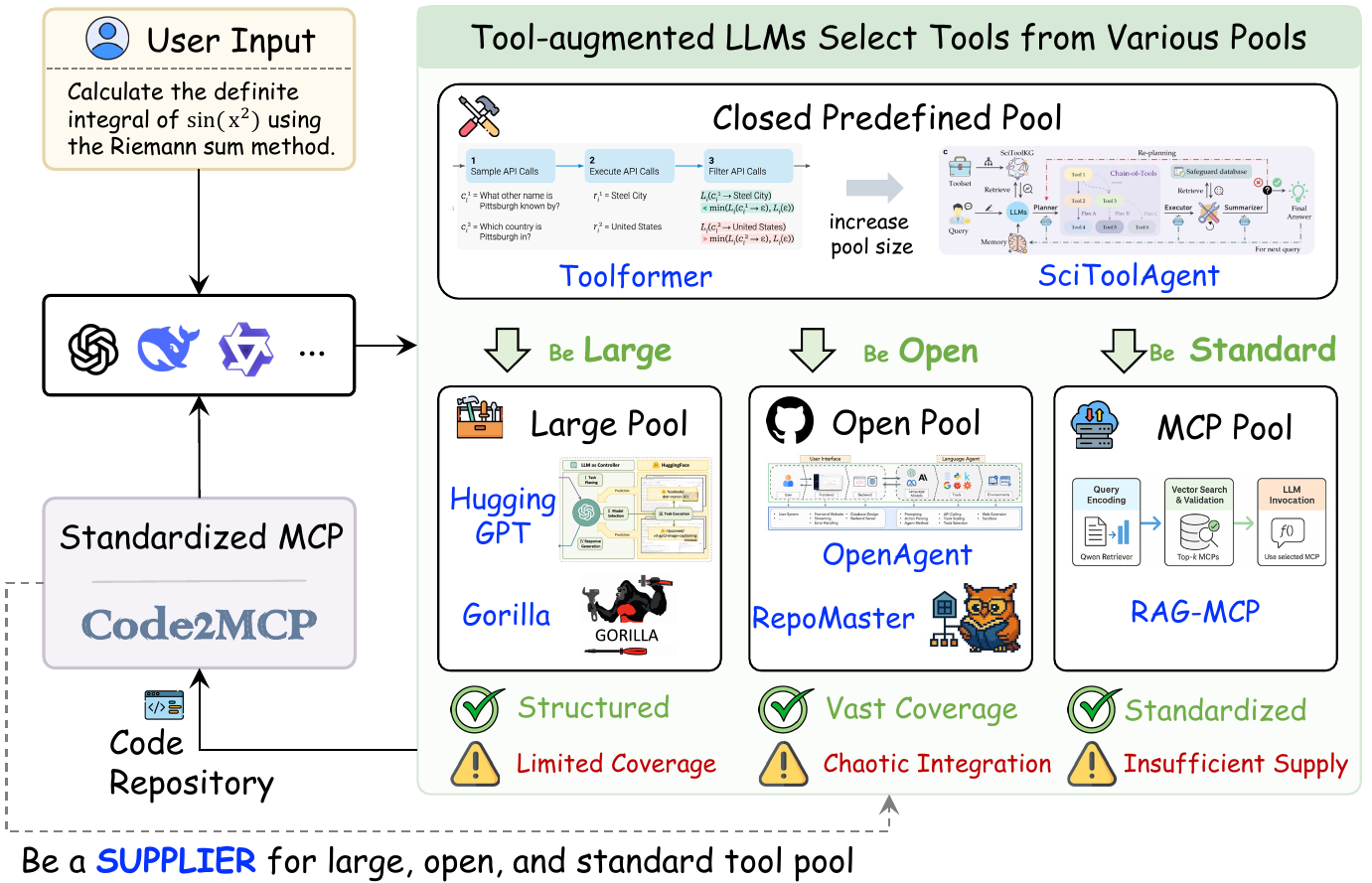}
  \caption{While most research focuses on the consumption of tools (right side), one bottleneck is their supply (left side). Code2MCP solves the supply problem by converting the code repository into a standardized MCP-compliant tool.}
  \label{fig:tool-use-overview}
\end{figure}
\begin{table*}[t]
\caption{A comparative summary of related works in tool-augmented LLMs.}
\label{tab:related_work_summary}
\centering
\setlength\tabcolsep{8pt}
\begin{tabular}{cllllc}
\toprule
\toprule
\bf Type & \textbf{Work} & \textbf{Core Contribution} & \textbf{Tool Pool} & \textbf{Tool Selection} & \textbf{MCP} \\ \midrule \addlinespace

\multirow{12}{*}{\rotatebox[origin=c]{90}{\textbf{Consumer}}} & Toolformer~\cite{schickToolformerLanguageModels2023}
& Teaching LLM to use external tools & 5 predefined tools & Fine-tuning & $\times$ \\ \addlinespace

& \multirow{2}{*}{SciToolAgent~\cite{ding2025scitoolagent}} & \multirow{2}{*}{\begin{tabular}[c]{@{}l@{}}Domain-specific enhancement \\ for scientific tool utilization \end{tabular}} & \multirow{2}{*}{\begin{tabular}[c]{@{}l@{}}KG of Scientific Tools\\(500+ tools)\end{tabular}}& \multirow{2}{*}{Retrieval on KG} & \multirow{2}{*}{$\times$} \\
& & & & & \\ \cmidrule{2-6} \addlinespace

& HuggingGPT~\cite{shen2023hugginggpt}
& \multirow{3}{*}{\begin{tabular}[c]{@{}l@{}}Increase the size\\of tool pool\end{tabular}} & \url{huggingface.co} & LLM task planning & $\times$ \\ \addlinespace

& \multirow{2}{*}{Gorilla~\cite{patilGorillaLargeLanguage2023}
}
& & \multirow{2}{*}{\begin{tabular}[c]{@{}l@{}}TorchHub, TensorHub\\ (1600+ tools)\end{tabular}} & \multirow{2}{*}{\begin{tabular}[c]{@{}l@{}}Retriever-aware\\ training\end{tabular}} & \multirow{2}{*}{$\times$} \\
& & & & & \\ \cmidrule{2-6} \addlinespace

& OpenAgents~\cite{OpenAgents}
& \multirow{2}{*}{\begin{tabular}[c]{@{}l@{}}Tool use from open-source \\ beyond a closed pool \end{tabular}} & \multirow{2}{*}{\url{github.com}} & Multi-agent planning & $\times$ \\ \addlinespace

& RepoMaster~\cite{wang2025repomaster}
& & & Rule-based deep search & $\times$ \\ \cmidrule{2-6} \addlinespace

& RAG-MCP~\cite{gan2025rag} & RAG for tool selection from MCP & \url{MCP.so}  & Retrieval on MCP & $\checkmark$ \\ \addlinespace \midrule

\bf Supplier & \bf Code2MCP & GitHub Repo to standardized MCP & \url{github.com} & N/A & $\checkmark$ \\

\bottomrule
\bottomrule
\end{tabular}
\end{table*}
In response to this integration challenge, the community's efforts have evolved, inadvertently revealing a deeper, more foundational bottleneck~\citep{yue2025autonomous}.
The initial challenge is to establish the fundamental feasibility of tool use, where a limited set of tools proves the feasibility of the paradigm~\citep{schickToolformerLanguageModels2023, patilGorillaLargeLanguage2023, ding2025scitoolagent,qin2024tool}. To break past the inherent scarcity of these platforms, the focus shifts to the vast landscape of open-source repositories~\citep{wang2025repomaster, OpenAgents,wang2023power}.

This move, however, trades a scarcity problem for a chaos problem, exposing the wild non-standardization of real-world code. The MCP emerges as a direct answer to this chaos, promising a universal interface~\citep{yu2025survey, anthropic2023mcp}. Yet, this leads to a critical gap: existing research focuses almost exclusively on the consumption side of MCP, using services from a presumed-to-exist pool~\citep{gan2025rag}, while the foundational supply-side problem of how to populate this pool from existing software remains largely unaddressed. We identify and formalize this largely overlooked tool supply problem in the MCP ecosystem: 
how to automatically transform existing code repositories into reusable, registrable, and agent-consumable MCP services, 
rather than relying on one-off, repository-level execution. As illustrated in Figure~\ref{fig:tool-use-overview}, existing approaches make different trade-offs among the 
large, open, and standard dimensions of tool pools,  
while the key bottleneck of the MCP pool lies in the insufficient supply of standardized tool services. 
Code2MCP is designed to act as a \emph{supplier} that fills this gap.

However, while these efforts advance the consumption side of the problem by improving how agents use tools, they largely overlook a more fundamental bottleneck on the supply side.
This bottleneck is not theoretical. Most consumer-side settings assume that a usable tool pool already exists, whether it is a curated knowledge graph as in SciToolAgent~\citep{ding2025scitoolagent} or open-source repositories as explored by OpenAgents~\citep{OpenAgents} and RepoMaster~\citep{wang2025repomaster}.
Yet the scale of available standardized tools remains limited.
For example, RAG-MCP~\citep{gan2025rag} performs retrieval over a pre-existing MCP server registry, with \url{mcp.so} as a representative platform, whereas the broader software ecosystem contains hundreds of millions of public GitHub repositories.
The critical question of how to create a large and diverse pool of these standardized, agent-ready tools has been left unaddressed.
This creates a major adoption gap, effectively locking away the largest software repository, GitHub, from this emerging ecosystem.
\begin{figure*}[!t]
  \centering  \includegraphics[width=\textwidth,height=0.28\textheight,keepaspectratio]{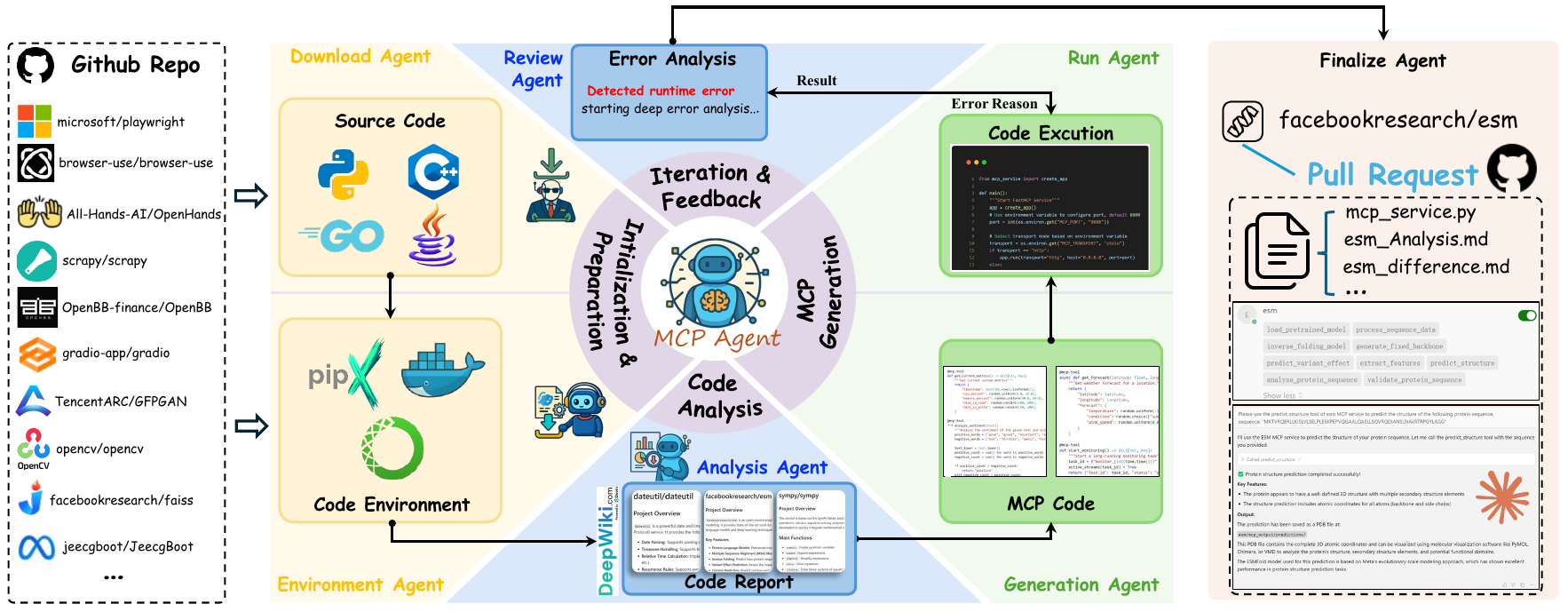}
  \caption{Overview of the Code2MCP framework. The system takes a GitHub repository URL as input and automatically generates a complete MCP service through a multi-agent workflow.}
  \label{fig:pipeline}
\end{figure*}
In this paper, we introduce Code2MCP, a new framework designed to bridge the critical tool supply gap. Code2MCP presents a blueprint for transforming a wide range of GitHub repositories into functional and documented MCP services with minimal human intervention. However, this transformation is a complex endeavor encompassing four pivotal challenges: (1) deep code comprehension to identify core functionalities, (2) reliable environment replication to ensure executability, (3) intelligent tool abstraction to design useful and valid service interfaces, and (4) robust self-correction to handle the inevitable errors throughout the process.
To systematically address these challenges, as shown in Figure~\ref{fig:pipeline}, Code2MCP implements a collaborative multi-agent system~\citep{park2023generative,yue2024clinicalagent,Tran2025MultiAgentCM,Guo2024LargeLM,dorri2018multi}. Unlike general-purpose coding agents~\citep{cognition_devin, jimenezSWEbenchCanLanguage2024,zhang2024codeagent}, different agents in our framework are specialized for the distinct stages of code analysis, environment setup, and API design. Crucially, the overall reliability of this workflow is ensured by an integrated Run-Review-Fix self-correction cycle, where failure traces are explicitly analyzed to generate targeted modification plans that constrain subsequent regeneration to specific files and code regions. This design differs from naive retry-based generation by enforcing localized, diagnosis-driven repairs rather than global re-synthesis, while keeping previously validated tool interfaces stable whenever possible.
The key contributions of this work are listed as follows:
\begin{itemize}[leftmargin=*]
\item To solve the fundamental tool supply bottleneck hindering the adoption of the MCP standard, we propose Code2MCP, a novel automated framework that, to the best of our knowledge, is the first to systematically transform code repositories into agent-ready MCP services.
\item The key challenge in converting code into a service is the inherent fragility of the multi-stage automation process, where an error at any step can derail the entire workflow. Thus, we introduce a novel multi-agent architecture governed by a Run-Review-Fix cycle, a self-correcting mechanism designed to systematically debug and refine the process, ensuring end-to-end reliability.
\item We demonstrate the effectiveness and scalability of our framework by converting highly complex and diverse scientific libraries, covering Protein Design, Symbolic Mathematics, and Computational Fluid Dynamics, into fully functional MCP services.
This provides a concrete and practical pathway to enrich the MCP ecosystem with specialized, high-value tools.
\end{itemize}

\section{Related Work}
\label{sec:related_work}

As summarized in Section~\ref{sec:intro} and Table~\ref{tab:related_work_summary}, the pioneering works focus on progressively expanding the scope of tool use, from initial feasibility studies using a few predefined APIs to leveraging large, curated tool platforms, and ultimately, to the ambitious goal of directly interfacing with unstructured open-source repositories.
Thus, the current bottleneck lies not in how LLMs consume tools, but in how such tools are supplied and created.
In this paper, Code2MCP is designed to solve this fundamental ``supply-side'' problem.

\textbf{Initial explorations in LLM Tool Use.}
The initial challenge is to establish the fundamental feasibility of tool use. Toolformer~\citep{schickToolformerLanguageModels2023} demonstrates that an LLM can learn to invoke simple, well-defined tools like a calculator via simple APIs in a self-supervised way.
SciToolAgent~\citep{ding2025scitoolagent} leverages knowledge graphs to orchestrate 500+ scientific tools.
This proves the concept and opens a new paradigm.
However, its reliance on a small, predefined set of tools is inherently unscalable and insufficient for addressing the diverse needs of real-world tasks.

\textbf{Scaling Tool Availability via Structured Platforms.}
To overcome the limitation of fixed toolsets, subsequent research turns to large, curated platforms. These approaches significantly expand the number of available tools. For instance, Gorilla~\citep{patilGorillaLargeLanguage2023} fine-tunes models on a massive corpus of API calls from hubs like TorchHub and TensorFlow Hub. Similarly, HuggingGPT~\citep{shen2023hugginggpt} positions an LLM as a controller to delegate tasks to specialized models within the Hugging Face ecosystem.
While powerful, their success hinges on environments where tools are well-documented and standardized.

\textbf{Exploring Unstructured Open-Source Repositories and Challenges.}
A more ambitious paradigm shift involves treating the entirety of open-source code repositories as a virtually infinite tool source. Frameworks like OpenAgents~\citep{OpenAgents} and RepoMaster~\citep{wang2025repomaster} empower agents to directly parse, reason about, and execute code within GitHub repositories. These works confront the complexity of real-world code but expose the core bottleneck: the vast majority of these repositories are not designed for programmatic use by LLM agents. They lack standardized interfaces~\citep{zhang2024large, jin2024llms, ray2025survey}, forcing the agent into an ad-hoc, brittle, and unreliable process of reverse-engineering the code, setting up its environment, and managing dependencies for every single task. While these systems partially touch the supply side by enabling agents to operate over open-source repositories, they do not aim to produce standardized, reusable MCP services that can be shared across agents and tasks.

\textbf{The Emergence of Standardization and Unaddressed Gap.}
Recognizing this chaos, the community has moved towards standardization,
exemplified by the MCP.
For example, RAG-MCP~\citep{gan2025rag} explores how an agent can effectively retrieve and select the most appropriate MCP service from a pool of available options.
This approach is promising, but it presumes the existence of a rich ecosystem of MCP-compliant services~\citep{hasan2025model}. This highlights a crucial gap: \emph{how is this ecosystem of MCP services populated in the first place?}

\section{Methodology: The Code2MCP Framework}
\label{sec:methodology}

To automatically transform an arbitrary GitHub repository into a fully functional and reliable MCP service, we design Code2MCP, an automated framework driven by the collaboration of seven specialized agents. The entire conversion process, as depicted in Figure~\ref{fig:pipeline}, is a multi-agent workflow that begins with code analysis, proceeds through a core Run-Review-Fix self-correction loop, and culminates in the generation of a merge-ready pull request. A detailed description of the framework is provided in Appendix~\ref{alg:mcp_agent}.

Suppose there exists a consumer-side work listed in Table~\ref{tab:related_work_summary}
that finds a suitable GitHub repository that may solve the user's query.
Code2MCP converts this repository into MCP that LLMs can call and use. This is the core difference between this ``supply-side'' work and the consumer-side works.

\textbf{Initialization and Analysis.}
The \texttt{Download Agent} first clones the specified repository, identified by its URL $u$, into an isolated local workspace.
The \texttt{Environment Agent} then replicates the runtime environment from dependency files or Dockerfiles, addressing one of the most common failure points in code conversion and supporting reliable subsequent code generation and testing.

Once the environment is ready, the \texttt{Analysis Agent} identifies tool-worthy functionalities within the codebase.
It leverages the \href{https://deepwiki.org/}{DeepWiki} tool to obtain a semantic view of the code and associates code entities with their intent from documentation and comments.
The output is a Code Report that summarizes candidate APIs and guides subsequent stages.

\textbf{Generation, Execution, and Self-Correction.}
Given this conversion blueprint, the framework enters its core iterative loop that turns the identified functionalities into executable MCP services.

The loop starts with the \texttt{Generation Agent}, which takes the Code Report and uses an LLM to abstract the core functionalities into MCP-compliant interfaces.
It creates the tool interface definitions and adapter file that connect the original code to the MCP interface, together with a basic test suite.

Once the code is generated, the \texttt{Run Agent} executes the test suite in the prepared environment to verify executability.
If the tests pass, the workflow proceeds to finalization; otherwise, the \texttt{Run Agent} records the error traceback $\tau$ and forwards it to the \texttt{Review Agent}.
The \texttt{Review Agent} analyzes $\tau$ together with the generated code, the Code Report, and the failing test, and diagnoses root causes such as logic errors, missing dependencies, or interface mismatches.
It then formulates a correction plan $\delta$ that specifies which files and code blocks to change, and hands this plan back to the Generation Agent to re-synthesize the MCP files.
This Run-Review-Fix loop repeats until tests pass or a maximum of $B$ attempts is reached.
\begin{figure}[!t]
  \centering
  \includegraphics[width=\linewidth]{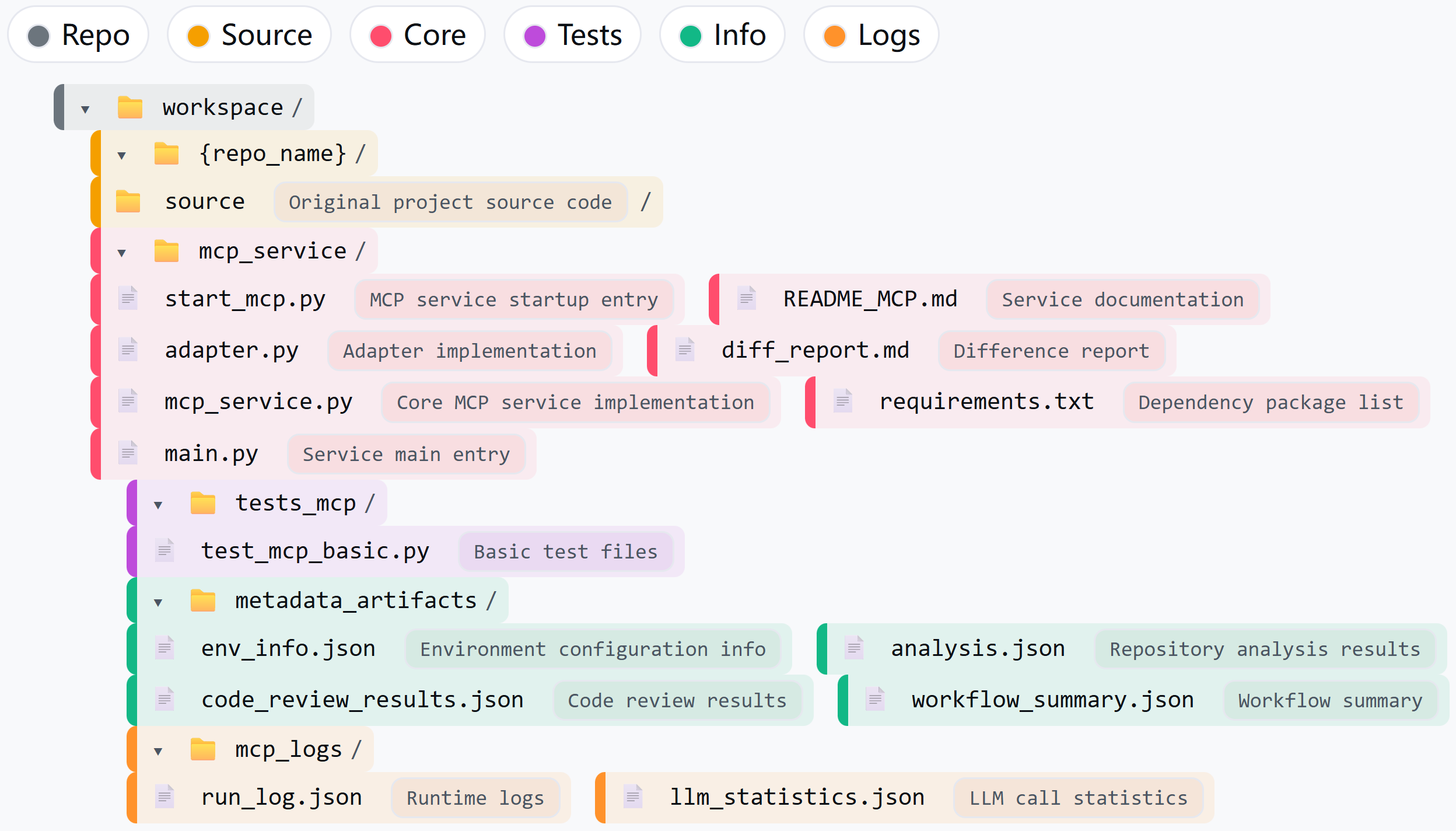}
  \caption{The complete output directory structure generated by the Code2MCP framework. The top-level \texttt{workspace} contains the original repository alongside all generated artifacts within the \texttt{mcp\_output} directory.}
  \label{fig:Output-Structure}
\end{figure}

\textbf{Finalization and Delivery.}
After the core loop succeeds, the \texttt{Finalize Agent} organizes and packages the validated MCP service files.
To facilitate review and adoption by the original repository maintainers, it also generates a README file explaining how to use the new MCP service. As shown in Figure~\ref{fig:Output-Structure}, All artifacts are arranged into a reproducible directory structure under the workspace, and the agent can prepare a pull request to submit these additions back to the original repository.

\textbf{Threat model and safety boundaries.}
Code2MCP executes third-party code during environment setup and validation, which poses inherent security risks.
In our evaluation, all conversions are run inside isolated worker environments with restricted filesystem scope and bounded runtime budgets.
By default, outbound network access is disabled during validation; for repositories whose core functionality intrinsically depends on external services, network access is selectively enabled under an explicit allowlist and treated as a separate evaluation condition.
We do not attempt to systematically evaluate adversarial repositories or malicious payloads in this work; instead, our security discussion focuses on operational safeguards during automated conversion rather than adversarial robustness.

\section{Evaluation}
\label{sec:evaluation}

\subsection{Experimental Setup}

\textbf{Task.}
The evaluation task is: given a GitHub repository, Code2MCP automatically generates an MCP service for that repository, and the outcome of the conversion is judged as a success or a failure.
The evaluation focuses on the overall conversion success rate across different domains and repository characteristics, and on the time and stability of Code2MCP compared with manual implementation and a GPT-4 template-based baseline on the same set of repositories.

\textbf{Repositories.}
Code2MCP is evaluated on 50 open-source repositories from 10 scientific and engineering domains (5 per domain), covering biomedical science, psychology, mathematical computing, earth science, chemistry, physics, astronomy, social science, linguistics, and econometrics.
The goal of this selection is not exhaustive coverage, but to expose the framework to a diverse range of dependency structures, API styles, and execution environments commonly encountered in scientific software.
For each repository, the evaluation logs the environment construction result, MCP conversion result, number of generated tools, and failure types observed across all Run-Review-Fix rounds; per-repository details and the associated failure labels are summarized in a large table in the Appendix~\ref{sec:appendix_repo_table}, and a subset of representative scientific-computing repositories is further used for the Run-Review-Fix ablation and qualitative case studies.

\textbf{Success criterion.}
A conversion is counted as successful iff all of the following hold:
(1) the runtime environment can be reconstructed from repository-provided specifications;
(2) the generated MCP server starts and passes the unified RunNode validation;
(3) at least three pre-identified, documentation-facing functionalities succeed under automatic invocation. These functionalities are selected deterministically from the repository's public documentation and matched to generated tools by name and semantic description, prior to conversion.

\textbf{Failure taxonomy.}
Each failed repository is annotated with one or more failure labels derived from execution artifacts (installation logs,  server logs, and schema validation).
If multiple error types occur across Run--Review--Fix rounds, labels are aggregated; a primary label is also recorded based on the final failing stage.
\begin{itemize}[leftmargin=*]
  \item \texttt{env\_failure}. The environment or dependencies cannot be reconstructed.
  \item \texttt{api\_inference\_error}. Systematic errors occur when inferring tool interfaces.
  \item \texttt{import\_error}. Import paths or cross-module dependencies are misconfigured.
  \item \texttt{repo\_internal\_bug}. Bugs or version conflicts inside the original repository prevent it from running reliably even in its native environment.
  \item \texttt{mcp\_spec\_violation}. The generated JSON Schema or response structure does not conform to the MCP specification.
  \item \texttt{untoolable\_repo}. The repository is a collection of scripts or heavily depends on interactive CLIs or GUIs, which are difficult to abstract into stable MCP tools.
\end{itemize}

\textbf{Implementation Details.}
By default, Code2MCP utilizes gpt-4o-2024-05-13 as its core reasoning engine.
The temperature for all models is set to 1.
Code2MCP leverages \texttt{gitingest}\footnote{\url{https://github.com/coderamp-labs/gitingest}} to ingest repositories into contextual prompts and fetch pre-analysis reports from \texttt{deepwiki}\footnote{\url{https://deepwiki.org}}.
Case studies are conducted on servers equipped with 8 NVIDIA H100 80 GB GPUs.

\begin{table*}[t]
\centering
\caption{Per-domain summary of environment setup success, basic test success, recovery by the Run-Review-Fix (RRF) loop (``--'' if none are recovered), and final MCP conversion success.}
\label{tab:domain_summary}
\setlength\tabcolsep{12pt}
\begin{tabular}{lcccccc}
\toprule
\toprule
\textbf{Domain} & \textbf{Repos} & \textbf{Env success} & \textbf{Test success} & \textbf{RRF recovered} & \textbf{Avg rounds} & \textbf{MCP success} \\
\midrule
\textbf{Biomedical}     & 5  & 4/5 (80\%)   & 2/5 (40\%)   & 1/2 (50\%)     & 1.5 & 3/5 (60\%)  \\
\textbf{Psychology}     & 5  & 3/5 (60\%)   & 3/5 (60\%)   & 0/1 (0\%)      & --  & 3/5 (60\%)  \\
\textbf{Math}           & 5  & 5/5 (100\%)  & 3/5 (60\%)   & 1/2 (50\%)     & 2.0 & 4/5 (80\%)  \\
\textbf{Earth Science}  & 5  & 3/5 (60\%)   & 2/5 (40\%)   & 1/3 (33.3\%)   & 1.0 & 3/5 (60\%)  \\
\textbf{Chemistry}      & 5  & 4/5 (80\%)   & 2/5 (40\%)   & 1/2 (50\%)     & 1.3 & 3/5 (60\%)  \\
\textbf{Physics}        & 5  & 3/5 (60\%)   & 2/5 (40\%)   & 0/2 (0\%)      & --  & 2/5 (40\%)  \\
\textbf{Astronomy}       & 5  & 4/5 (80\%)   & 3/5 (60\%)   & 1/1 (100\%)    & 1.5 & 4/5 (80\%)  \\
\textbf{Social Science} & 5  & 4/5 (80\%)   & 2/5 (40\%)   & 1/2 (50\%)     & 1.0 & 3/5 (60\%)  \\
\textbf{Linguistics}    & 5  & 5/5 (100\%)  & 3/5 (60\%)   & 1/2 (50\%)     & 2.0 & 4/5 (80\%)  \\
\textbf{Econometrics}   & 5  & 3/5 (60\%)   & 2/5 (40\%)   & 1/2 (50\%)     & 1.0 & 3/5 (60\%)  \\
\midrule
\textbf{Overall}        & 50 & 38/50 (76\%) & 24/50 (48\%) & 8/19 (42.1\%)  & 1.4 & 32/50 (64\%) \\
\bottomrule
\bottomrule
\end{tabular}
\end{table*}

\subsection{Large-scale Repository Conversion and Failure Analysis}

\textbf{Overall and per-domain success.}
The overall conversion success rate is first analyzed at the domain level.
\begin{figure*}[htbp]
\centering
\begin{subfigure}[t]{0.48\linewidth}
\centering
\includegraphics[width=\linewidth]{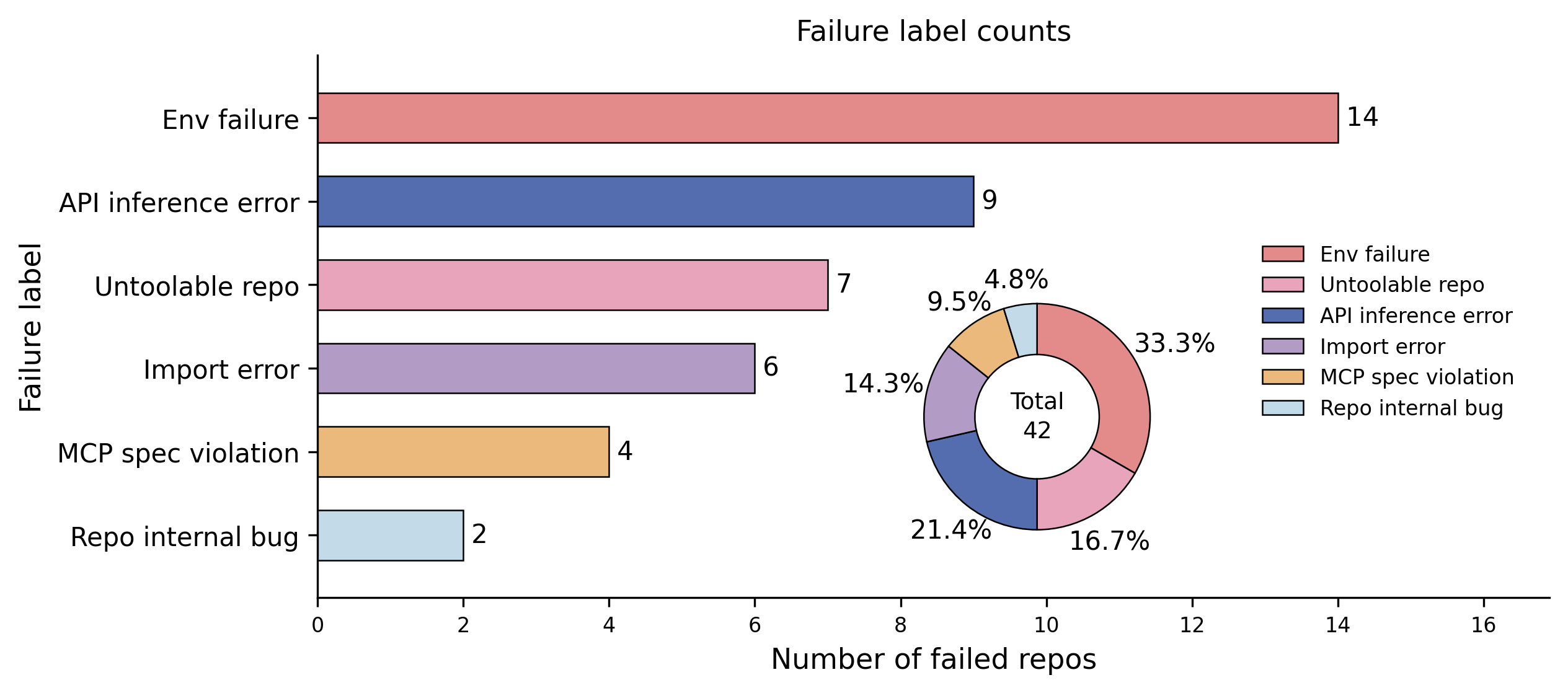}
\caption{Failure label distribution.}
\label{fig:exp1_failure_summary}
\end{subfigure}\hspace{0.02\linewidth}
\begin{subfigure}[t]{0.48\linewidth}
\centering
\includegraphics[width=\linewidth]{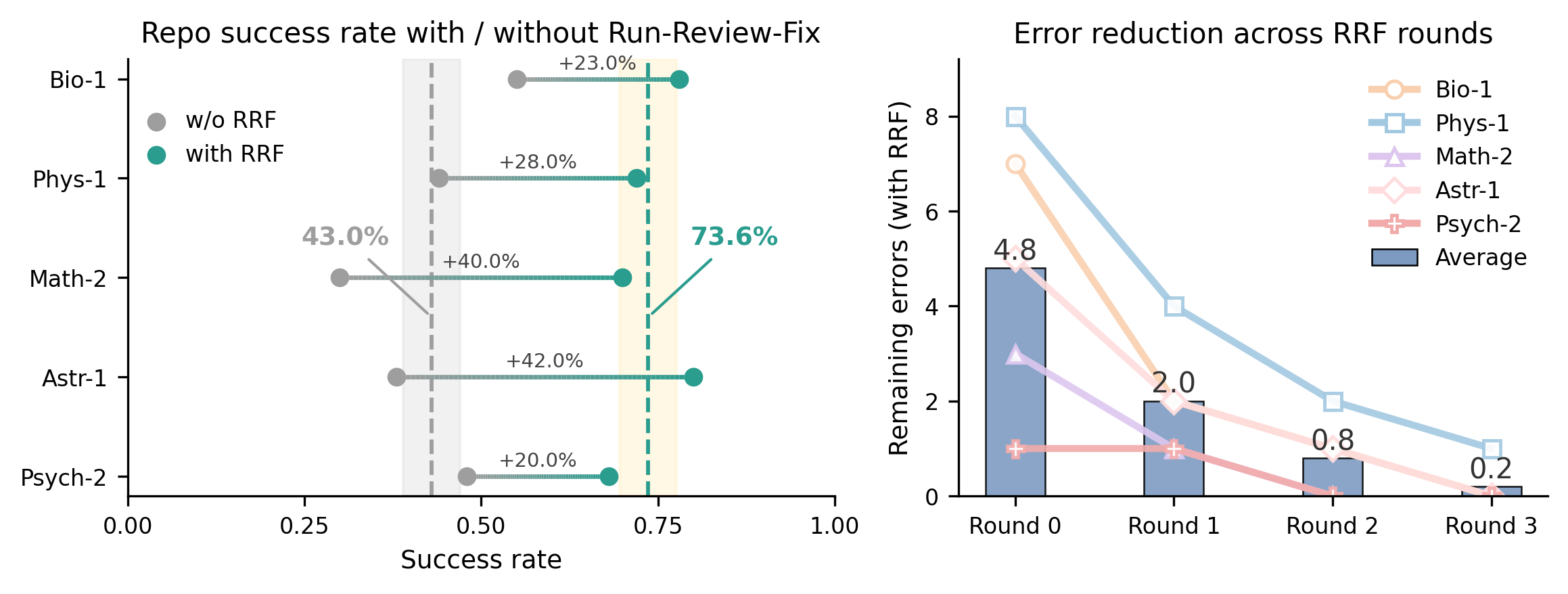}
\caption{RRF ablation results.}
\label{fig:exp3_rrf_ablation}
\end{subfigure}
\label{fig:exp1_exp3_combined}
\caption{(a) summarizes the distribution of 42 failure labels assigned to 18 failed repositories; a single repository can trigger multiple failure types across different Run-Review-Fix rounds. (b) reports repository success rates with and without Run‑Review‑Fix and the remaining errors across RRF rounds for five representative scientific MCP repositories.}
\end{figure*}
In Table~\ref{tab:domain_summary}, for each domain, the number of repositories with successful environment setup and basic test passing, the number of initially failing repositories recovered by the Run-Review-Fix loop and their average rounds, and the final number of successful MCP conversions.
The per-domain MCP success rates range from 0.40 to 0.80:
domains dominated by library-style projects, such as mathematical computing, astronomy, and linguistics, tend to have higher success rates,
while domains such as physics and econometrics, which contain more complex environments and mixtures of scripts and workflow-style code, tend to have lower success rates.
All statistics in the table are computed on a per-repository basis, where each repository contributes a single final outcome after the Run-Review-Fix loop.
Per-repository results, including environment construction and MCP conversion outcomes as well as the number of generated tools, are summarized in the large table in the appendix and can be further grouped by domain or repository size if desired.

\textbf{Failure modes.}
To understand why conversions fail on real repositories, all 18 failed repositories in the 50-repository evaluation are annotated using the six failure labels introduced in the failure taxonomy.
Labels are aggregated across all Run-Review-Fix rounds, so a single repository may receive multiple labels if different error types are observed during different stages of the pipeline.
The distribution of these labels is visualized in Figure~\ref{fig:exp1_failure_summary}.

The labels \texttt{env\_failure} and \texttt{api\_inference\_error} account for more than half of all failures, indicating that environment reconstruction and interface inference are the primary bottlenecks of the current workflow.
The labels \texttt{import\_error} and \texttt{untoolable\_repo} are also relatively common, reflecting the difficulty of handling complex import paths, initialization order, and script-driven repositories in a fully automated pipeline.
\begin{figure*}[!t]
\centering
\includegraphics[width=\linewidth]{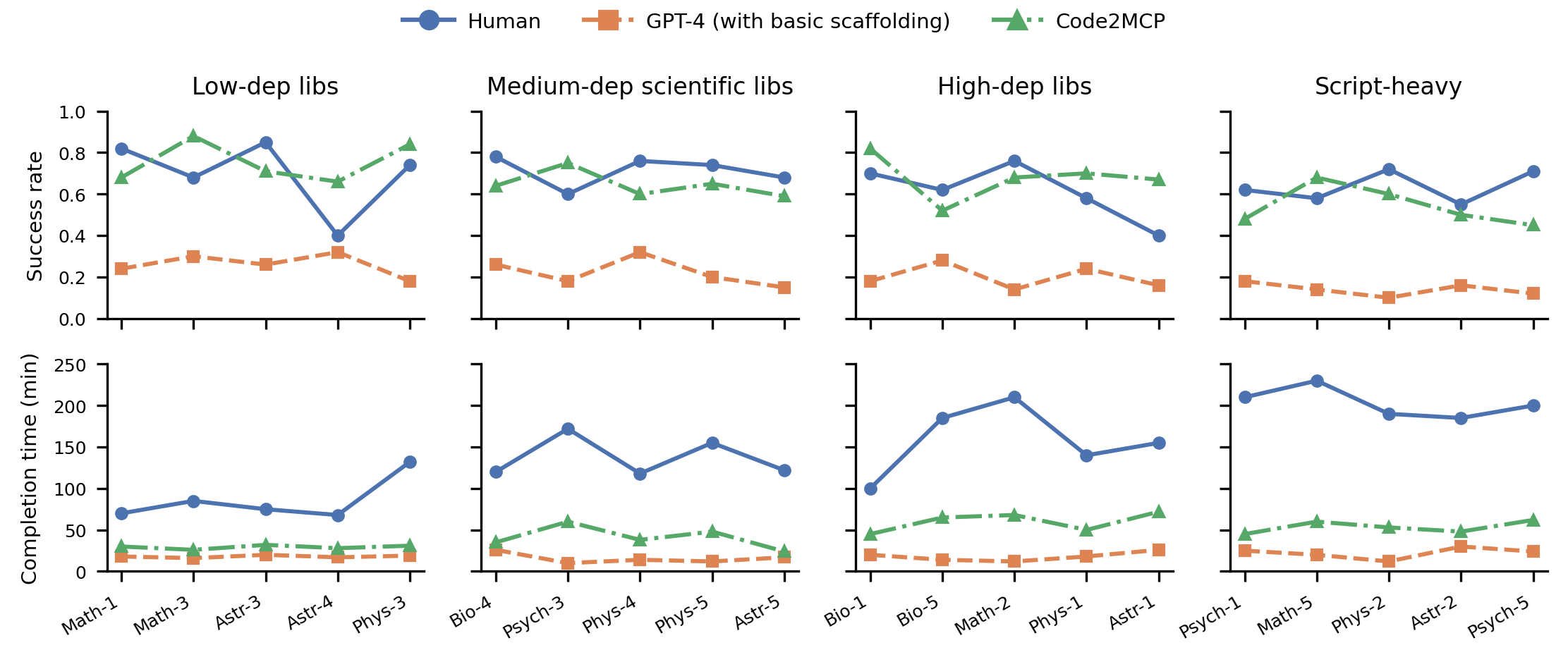}
\caption{
Task success rates (top row) and completion time (bottom row) for Human experts, GPT-4 with a template, and Code2MCP across representative repositories in four repository groups. The x-axis lists representative repositories in each group, labeled by domain abbreviation and index (e.g., Math-1).}
\label{fig:exp2_success_time_by_bucket}
\end{figure*}

\textbf{Run-Review-Fix ablation.}
To assess the specific contribution of the Run-Review-Fix self-correction loop, an ablation study is conducted on a set of representative scientific-computing repositories by comparing two configurations: a single-pass configuration without the loop and the full Code2MCP pipeline with the loop enabled.
Figure~\ref{fig:exp3_rrf_ablation} summarizes the results.
The bar chart on the left shows that the per-repository success rates increase for most repositories once the loop is enabled, often moving from medium success under a single-generation setting to success rates close to or on par with human-written wrappers.
The line chart on the right plots the average number of remaining errors as a function of the number of Run-Review-Fix rounds, where ``remaining errors'' denote the number of failing assertions or uncaught exceptions in the unified test after each round.
These results indicate the majority of initial failures can be automatically repaired within one to three Run-Review-Fix rounds, whereas the few repositories that still fail after multiple rounds represent the main  limitation of the method.

\subsection{Comparison with Human and GPT-4 Baselines}

The evaluation compares three configurations on the same set of representative repositories:
\begin{itemize}[leftmargin=*]
  \item a human configuration, where MCP wrappers are implemented from scratch by human developers;
  \item a GPT-4 template configuration, where a single GPT-4 agent generates MCP wrappers;
  \item the Code2MCP configuration, which uses the multi-agent pipeline with the Run-Review-Fix loop.
\end{itemize}

In the human configuration, ten graduate students with at least three years of Python experience and basic familiarity with MCP concepts implement wrappers from scratch. Before the study, all participants read the MCP specification and a commented example repository. Each participant is randomly assigned several repositories.
For consistency, the same success criterion and the same validation procedure are applied. Completion time is measured from the start of wrapper implementation to the first time the unified validation passes, or to termination under the same budgets.

In the GPT-4 template configuration, a unified system prompt instructs a single model to generate a complete MCP service implementation based on the repository README and several key source files selected by static analysis. The model performs end-to-end interface design, code generation, execution, and iterative debugging based on execution tracebacks, constituting a single-agent baseline.
The baseline follows the same validation procedure and the same success criterion. Generation and revision are performed iteratively based on execution errors returned by the test runner, and the process stops when the unified validation passes or when the evaluation budget is exhausted. No additional agents, retrieval modules, or external planning tools are used. The full system prompt is provided in Appendix~\ref{sec:appendix_gpt4_template_prompt}.
\begin{figure*}[!t]
  \centering
  \begin{subfigure}[t]{0.48\linewidth}
    \centering
    \includegraphics[width=\linewidth]{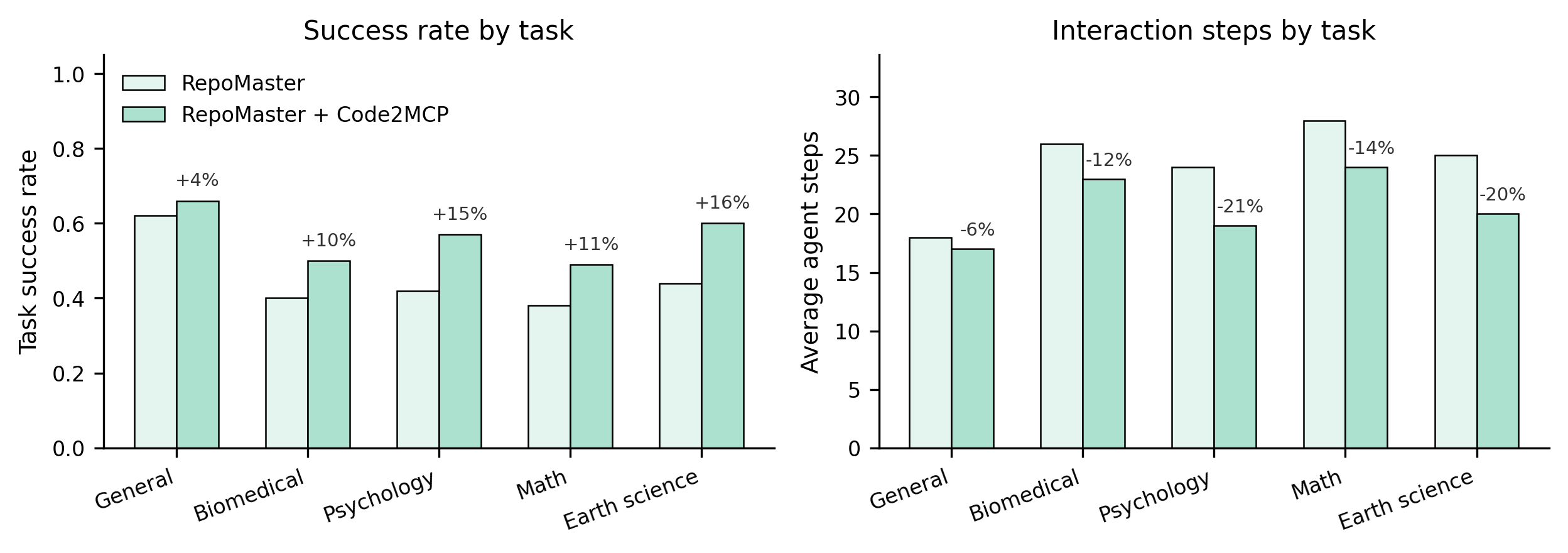}
    \caption{RepoMaster with/without Code2MCP.}
    \label{fig:exp4_repomaster}
  \end{subfigure}\hspace{0.02\linewidth}
  \begin{subfigure}[t]{0.48\linewidth}
    \centering
    \includegraphics[width=\linewidth]{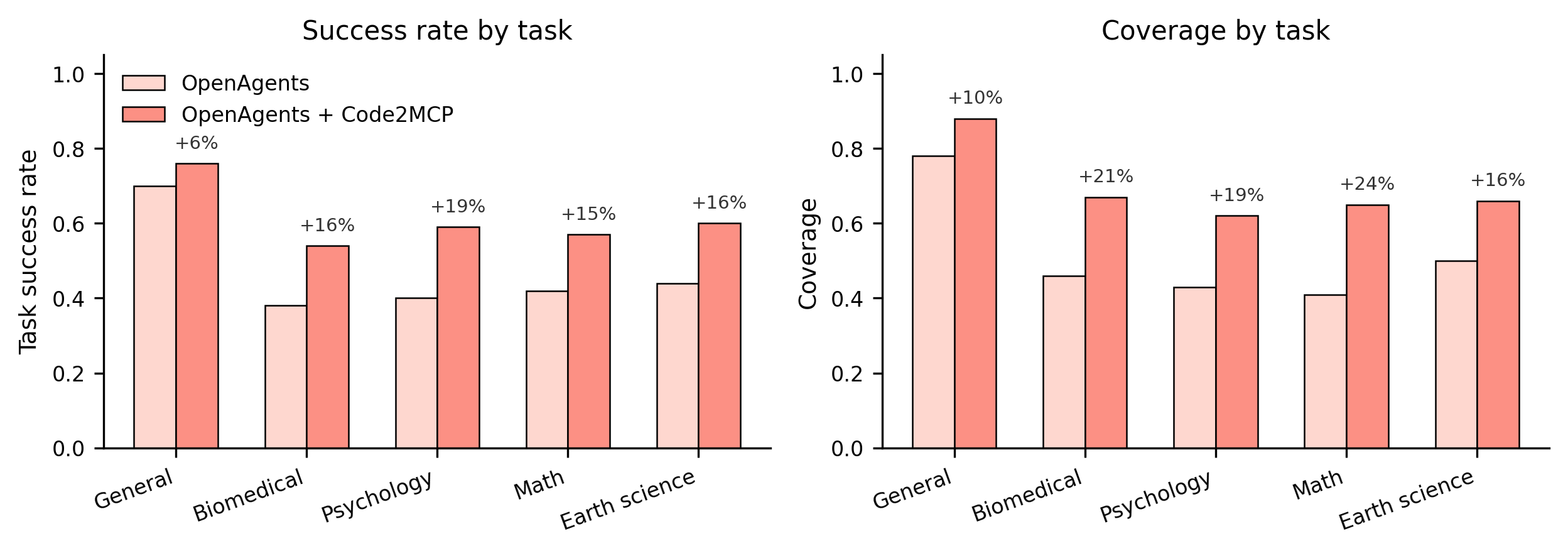}
    \caption{OpenAgents with/without Code2MCP.}
    \label{fig:exp4_openagents}
  \end{subfigure}
\caption{(a) shows RepoMaster with and without Code2MCP tools, reporting task success rates and average interaction steps across task groups. (b) shows OpenAgents with and without Code2MCP tools, reporting task success rates and coverage across general and scientific task groups.}
  \label{fig:exp4_combined}
\end{figure*}

In the Code2MCP configuration, the default multi-agent pipeline with the Run-Review-Fix loop is used.
For each repository, we record the final success status and the completion time, defined as the end-to-end wall-clock time starting from the beginning of environment setup and ending when the unified MCP test first passes or the run is terminated. To make the evaluation protocol fully self-contained.

\begin{figure}[!h]
  \centering
  \includegraphics[width=\linewidth]{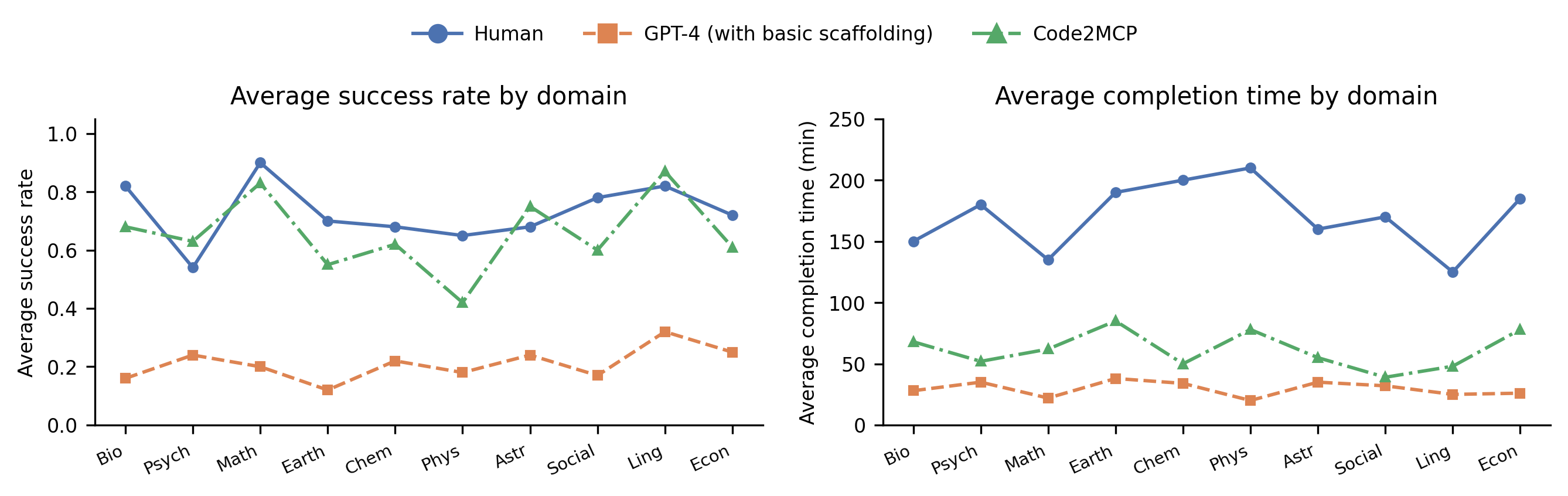}
  \caption{Average task success rate (left) and average completion time (right) across the ten scientific domains for the three configurations: Human experts, GPT-4 with basic scaffolding, and Code2MCP.}
  \label{fig:exp2_method_comparison}
\end{figure}

Figure~\ref{fig:exp2_method_comparison} summarizes the average task success rates and completion times of the three configurations across the ten scientific domains.
For each configuration and domain, task success rate is defined as the fraction of successful attempts over all attempts on representative repositories in that domain.
Completion time is the average end-to-end wall-clock time per attempt, measured from the beginning of environment setup until the minimal MCP test first passes or the run is terminated.
The overall trend is as follows:
Code2MCP achieves task success rates close to human implementations in most domains, and higher than the GPT-4 template configuration.
In terms of completion time, human implementations are on the order of hours, while Code2MCP completes in tens of minutes; the GPT-4 template configuration is the fastest but has lower success rates.
When repositories are grouped by dependency complexity and project type into low-dependency libraries, medium-dependency scientific libraries, high-dependency libraries, and script-heavy projects, Figures~\ref{fig:exp2_success_time_by_bucket} further break down the same comparison on representative repositories within each group.

\subsection{Integration with Existing Tool Systems}
\label{sec:integration}

This subsection summarizes how Code2MCP integrates with existing tool ecosystems by combining it with RepoMaster and OpenAgents on a shared set of tool-usage tasks.

\textbf{Task Set Description.}
In the integration experiments with RepoMaster and OpenAgents, we evaluate on a fixed set of code-understanding and tool-usage tasks designed to assess whether automatically generated MCP tools can be discovered, selected, and invoked by existing agent systems under realistic usage patterns.
The task categories and evaluation protocol are kept identical across methods, and full task specifications and examples are provided in Appendix~\ref{sec:appendix_tasks}.

\textbf{RepoMaster + Code2MCP.}
In the RepoMaster setting, tool-usage and code-understanding tasks are run in two configurations: using RepoMaster alone to operate at the source-code level, and using RepoMaster together with Code2MCP, where RepoMaster first selects relevant repositories and then invokes the corresponding MCP tools.
Figure~\ref{fig:exp4_repomaster} shows that the configuration with Code2MCP achieves higher task success rates and requires fewer interaction steps, suggesting that delegating part of the work to standardized MCP tools improves efficiency.

\textbf{OpenAgents + Code2MCP.}
In the OpenAgents setting, cross-domain tasks are evaluated under two configurations: using OpenAgents alone, and using OpenAgents with MCP tools automatically generated by Code2MCP added to the tool pool. A task is considered successful if the agent produces a final answer that passes automatic correctness checks.

\begin{figure*}[t]
  \centering
  \begin{subfigure}[t]{0.48\linewidth}
    \centering
    \includegraphics[width=\linewidth]{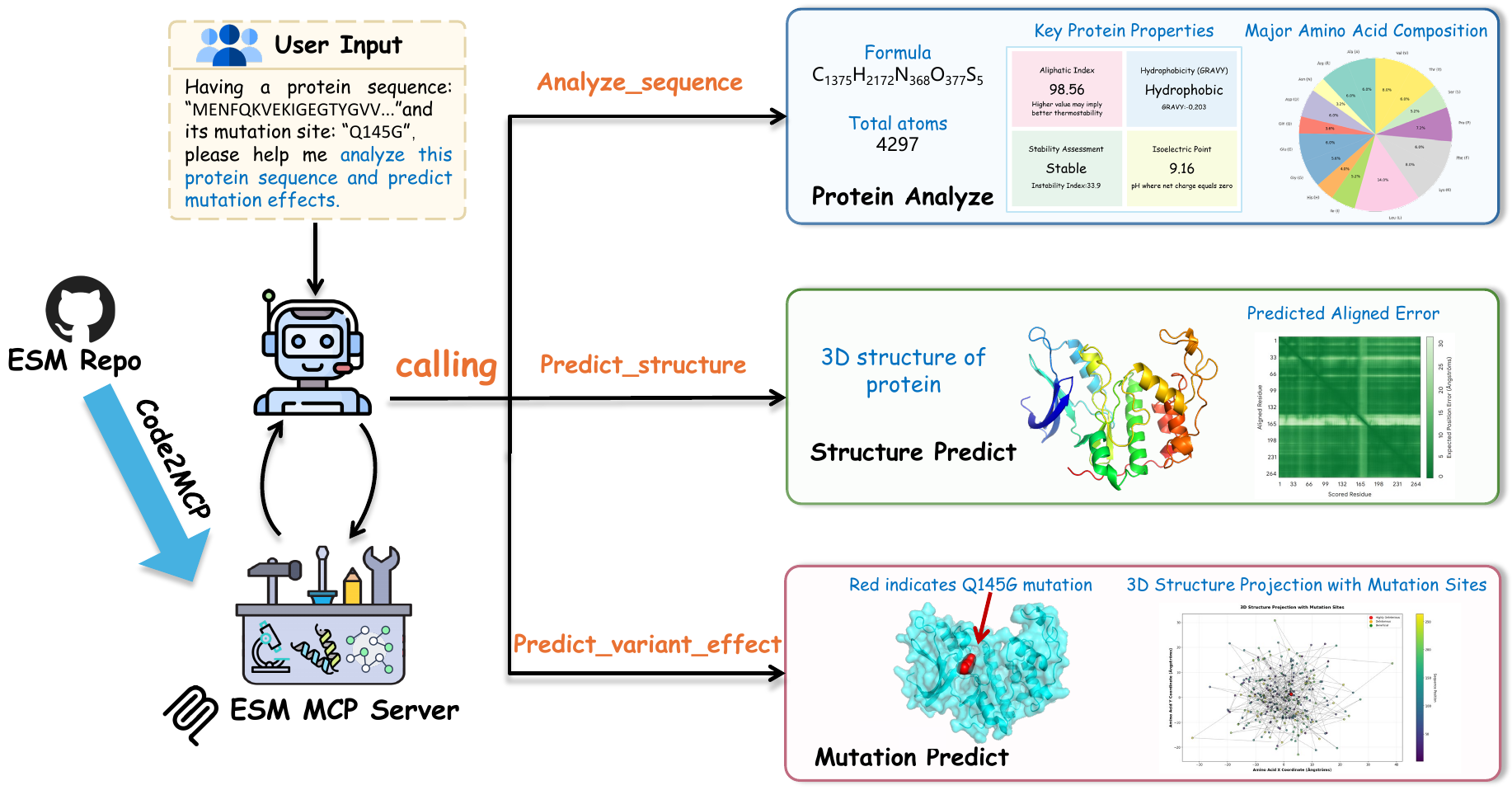}
    \caption{ESM protein analysis workflow.}
    \label{fig:case_study_esm}
  \end{subfigure}
  \hspace{0.02\linewidth}
  \begin{subfigure}[t]{0.48\linewidth}
    \centering
    \includegraphics[width=\linewidth]{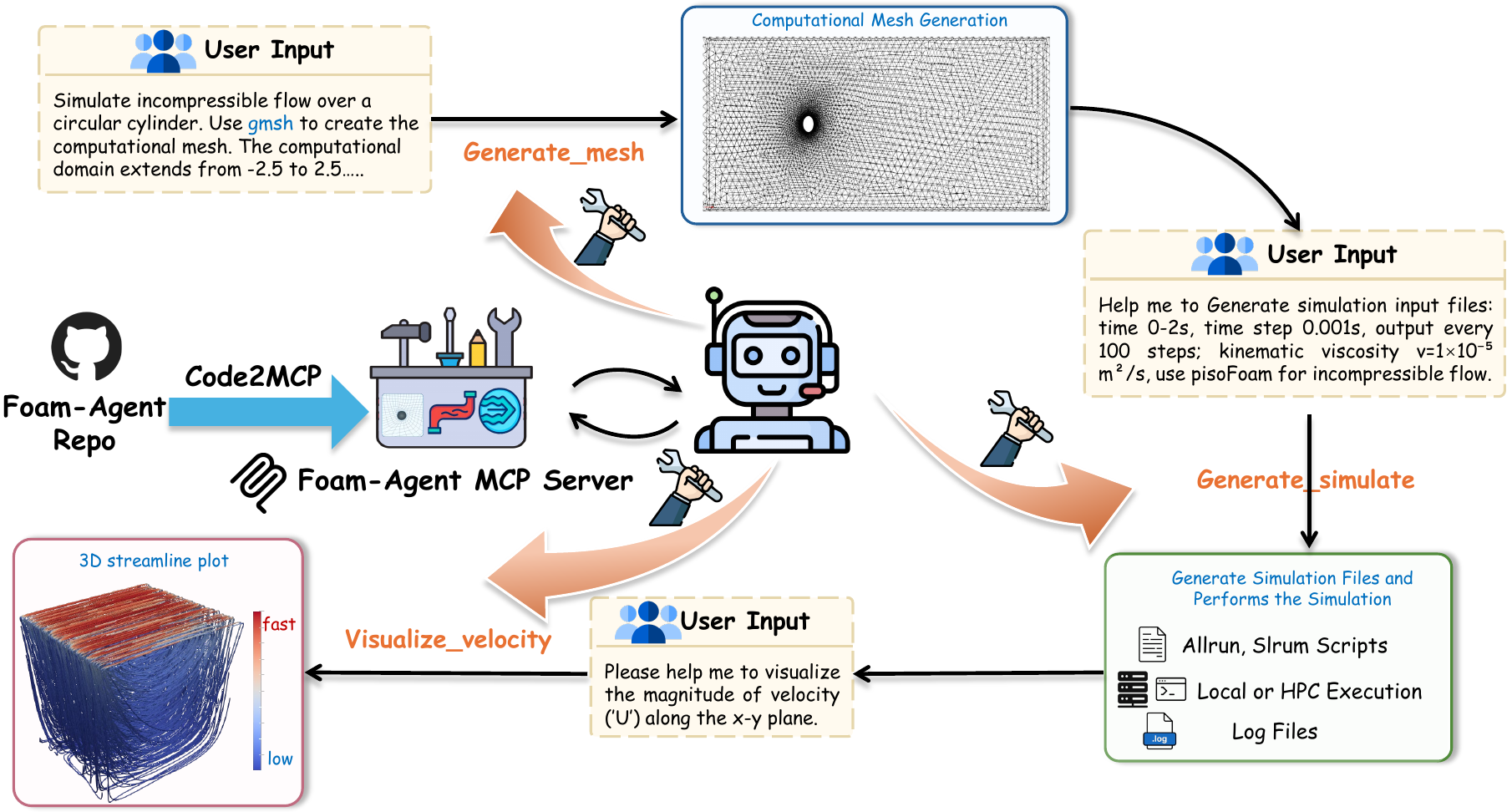}
    \caption{Foam-Agent CFD workflow.}
    \label{fig:case_study_foamagent}
  \end{subfigure}
\caption{(a) shows a protein analysis pipeline where an agent invokes \texttt{Analyze\_sequence}, \texttt{Predict\_structure}, and \texttt{Predict\_variant\_effect} from the generated \texttt{ESM} MCP server to return physicochemical properties, a predicted 3D structure, and mutation impact analysis.  (b) shows a CFD pipeline where the agent calls \texttt{Generate\_mesh}, \texttt{Generate\_simulate}, and \texttt{Visualize\_velocity} to guide the user from mesh generation and solver setup through simulation execution and final velocity-field visualization.}
  \label{fig:case_study_combined}
\end{figure*}

Task success rate is the fraction of tasks in a group that are successful.
Coverage is the fraction of tasks in a group for which the agent produces any final answer that can be checked by the automatic evaluator, regardless of whether the answer is ultimately judged correct.
As shown in Figure~\ref{fig:exp4_openagents}, we report the task success rate and coverage.
Adding Code2MCP tools increases both metrics across all task groups, indicating that a richer MCP tool pool can be effectively exploited by existing planning and retrieval strategies.

Overall, RepoMaster and OpenAgents handle tool discovery and code retrieval on the consumer side, while Code2MCP supplies additional MCP services on the supply side.
Under the same front-end selection and planning strategies, integrating Code2MCP enriches the tool pool and yields measurable improvements.

\subsection{Case Studies: Protein, Math, and Computational Fluid Dynamics}

This subsection highlights three representative repositories from the 50-repository evaluation set: the biomedical protein modeling library ESM, the symbolic mathematics library SymPy, and the CFD framework Foam-Agent~\citep{yue2025foamv2} built on OpenFOAM, illustrating the kinds of MCP tools Code2MCP generates and how agents use them in practice.

\textbf{Bridging protein science with AI agents.}
In protein science, models such as AlphaFold~\citep{jumper2021highly} have greatly improved structure prediction, but using them in everyday research still often requires substantial scripting and environment setup. ESM models complement AlphaFold with efficient sequence modeling and zero-shot variant effect prediction, yet many common analyses remain multi-step Python pipelines for computing physicochemical properties, predicting structures, and interpreting mutations. This workflow can be distracting for researchers who primarily focus on wet-lab experiments and want results that are easy to reproduce and explain.
Code2MCP converts ESM into an MCP service by exposing tools such as \texttt{Analyze\_sequence}, \texttt{Predict\_structure}, and \texttt{Predict\_variant\_effect}. Figure~\ref{fig:case_study_esm} illustrates a qualitative example. When a user provides a protein sequence and a mutation site like ``Q145G'', the agent can first call \texttt{Analyze\_sequence} to produce a baseline characterization, including molecular weight, amino acid composition, and hydrophobicity. It then calls \texttt{Predict\_structure} to generate a 3D structure and uses a Predicted Aligned Error (PAE) plot to assess reliability. Finally, \texttt{Predict\_variant\_effect} helps locate the mutation on the structure and summarize likely impacts, and it can also generate a simple visualization to clarify the mutation's spatial context.

\textbf{Enhancing mathematical reliability in AI agents.}
For mathematical computing, large language models are prone to errors in symbolic derivations and exact calculations, whereas libraries such as SymPy already provide reliable implementations. Code2MCP organizes common capabilities from SymPy into a mathematical MCP service with tools for limits and integration, matrix operations, symbolic simplification, and transforms such as Fourier and Laplace. For example, to find the indefinite integral of \(y=\ln(x^2+1)\), it calls the corresponding integration tool, and for the volume of revolution of \(y=\ln(x)\cdot\sin(x)\) from \(x=1\) to \(x=2\pi\), it invokes the volume tool to produce the exact result. By grounding sensitive mathematical steps in validated code rather than prompt-based manipulation, the workflow becomes easier to verify and reproduce.

\textbf{Automating CFD simulation for AI agents.}
CFD workflows typically involve mesh generation, solver configuration, execution, and visualization, traditionally coordinated by engineers through shell scripts and configuration files—an error-prone and labor-intensive process. Foam-Agent provides a higher-level interface to OpenFOAM, and Code2MCP further turns it into an MCP service with tools such as \texttt{Generate\_mesh}, \texttt{Generate\_simulate}, and \texttt{Visualize\_velocity}, allowing an agent to run the workflow step-by-step in dialogue. As illustrated in Figure~\ref{fig:case_study_foamagent}, a user describes the target geometry and physical parameters in natural language, and the agent calls these tools in sequence to construct the mesh, run the simulation, and visualize the velocity field.  

\section{Conclusion}
This paper addresses the key challenge of insufficient tool supply in the MCP ecosystem by introducing Code2MCP, a framework that automatically converts GitHub repositories into functional MCP services. The framework employs a multi-agent workflow for code comprehension, environment reconstruction, and tool abstraction, augmented by a ``Run-Review-Fix'' self-correcting loop that improves end-to-end reliability across diverse scientific libraries. An evaluation on 50 repositories demonstrates that Code2MCP can robustly expose high-value APIs as MCP tools while also revealing clear failure modes and boundary conditions. Overall, the results highlight an automated pathway for connecting existing codebases to the agent tool ecosystem, while surfacing concrete limitations that delimit the feasibility of fully automated tool supply.

\clearpage
\bibliographystyle{ACM-Reference-Format}
\bibliography{Code2MCP}

@article{qin2024tool,
  title={Tool Learning with Foundation Models},
  author={Qin, Yujia and Hu, Shengding and Lin, Yankai and Chen, Weize and Ding, Ning and Cui, Ganqu and Zeng, Zheni and Zhou, Xuanhe and Huang, Yufei and Xiao, Chaojun and Han, Chi and Fung, Yi Ren and Su, Yusheng and Wang, Huadong and Qian, Cheng and Tian, Runchu and Zhu, Kunlun and Liang, Shihao and Shen, Xingyu via and Xu, Bokai and Zhang, Zhen and Ye, Yining and Li, Bowen and Tang, Ziwei and Yi, Jing and Zhu, Yuzhang and Dai, Zhenning and Yan, Lan and Cong, Xin and Lu, Yaxi and Zhao, Weilin and Huang, Yuxiang and Yan, Junxi and Han, Xu and Sun, Xian and Li, Dahai and Phang, Jason and Yang, Cheng and Wu, Tongshuang and Ji, Heng and Li, Guoliang and Liu, Zhiyuan and Sun, Maosong},
  journal={ACM Computing Surveys},
  volume={57},
  number={4},
  pages={1--40},
  year={2024},
  publisher={ACM New York, NY, USA},
  doi={10.1145/3704435}
}

@article{wang2023power,
  title={The power of openness-how open source software is reshaping software engineering and industrial adoption},
  author={Wang, Yifei},
  journal={Authorea Preprints},
  year={2023},
  publisher={Authorea}
}

@inproceedings{yue2024clinicalagent,
  title={Clinicalagent: Clinical trial multi-agent system with large language model-based reasoning},
  author={Yue, Ling and Xing, Sixue and Chen, Jintai and Fu, Tianfan},
  booktitle={Proceedings of the 15th ACM International Conference on Bioinformatics, Computational Biology and Health Informatics},
  pages={1--10},
  year={2024}
}

@article{Tran2025MultiAgentCM,
  title={Multi-Agent Collaboration Mechanisms: A Survey of LLMs},
  author={Khanh-Tung Tran and Dung Dao and Minh-Duong Nguyen and Quoc-Viet Pham and Barry O’Sullivan and Hoang D. Nguyen},
  journal={ArXiv},
  year={2025},
  volume={abs/2501.06322},
  url={https://api.semanticscholar.org/CorpusID:275471465}
}

@inproceedings{Guo2024LargeLM,
  title={Large Language Model based Multi-Agents: A Survey of Progress and Challenges},
  author={Taicheng Guo and Xiuying Chen and Yaqi Wang and Ruidi Chang and Shichao Pei and N. Chawla and Olaf Wiest and Xiangliang Zhang},
  booktitle={International Joint Conference on Artificial Intelligence},
  year={2024},
  url={https://api.semanticscholar.org/CorpusID:267412980}
}

@article{dorri2018multi,
  title={Multi-agent systems: A survey},
  author={Dorri, Ali and Kanhere, Salil S and Jurdak, Raja},
  journal={Ieee Access},
  volume={6},
  pages={28573--28593},
  year={2018},
  publisher={IEEE}
}

@inproceedings{zhang2024codeagent,
    title = "{C}ode{A}gent: Enhancing Code Generation with Tool-Integrated Agent Systems for Real-World Repo-level Coding Challenges",
    author = "Zhang, Kechi  and
      Li, Jia  and
      Li, Ge  and
      Shi, Xianjie  and
      Jin, Zhi",
    editor = "Ku, Lun-Wei  and
      Martins, Andre  and
      Srikumar, Vivek",
    booktitle = "Proceedings of the 62nd Annual Meeting of the Association for Computational Linguistics (Volume 1: Long Papers)",
    month = aug,
    year = "2024",
    address = "Bangkok, Thailand",
    publisher = "Association for Computational Linguistics",
    url = "https://aclanthology.org/2024.acl-long.737/",
    doi = "10.18653/v1/2024.acl-long.737",
    pages = "13643--13658"
}

@misc{cognition_devin,
  title = {Devin: The First AI Software Engineer},
  author = {Cognition},
  year = {2024},
  url = {https://www.cognition-labs.com/},
  urldate = {2025-08-28}
}

@misc{parisiTALMToolAugmented2022,
  title = {TALM: Tool Augmented Language Models},
  author = {Parisi, Aaron and Zhao, Yao and Fiedel, Noah},
  year = {2022},
  month = may,
  number = {arXiv:2205.12255},
  eprint = {2205.12255},
  primaryclass = {cs},
  publisher = {arXiv},
  doi = {10.48550/arXiv.2205.12255},
  urldate = {2025-08-28},
  archiveprefix = {arXiv}
}

@techreport{anthropic2023mcp,
  title = {Model Context Protocol},
  author = {Anthropic},
  year = {2023},
  month = nov,
  institution = {Anthropic},
  url = {https://modelcontextprotocol.io/},
  urldate = {2025-08-28}
}

@article{park2023generative,
  title={Generative agents: Interactive simulacra of human behavior},
  author={Park, Joon Sung and O'Brien, Joseph C and Cai, Carrie J and Morris, Meredith Ringel and Liang, Percy and Bernstein, Michael S},
  journal={arXiv preprint arXiv:2304.03442},
  year={2023}
}

@inproceedings{hao2023reasoning,
  title={Reasoning with language model is planning with world model},
  author={Hao, Shibo and Gu, Yi and Ma, Haodi and Hong, Joshua and Wang, Zhen and Wang, Daisy and Hu, Zhiting},
  booktitle={Proceedings of the 2023 Conference on Empirical Methods in Natural Language Processing},
  pages={8154--8173},
  year={2023}
}

@inproceedings{yu2025survey,
  title={A survey on trustworthy llm agents: Threats and countermeasures},
  author={Yu, Miao and Meng, Fanci and Zhou, Xinyun and Wang, Shilong and Mao, Junyuan and Pan, Linsey and Chen, Tianlong and Wang, Kun and Li, Xinfeng and Zhang, Yongfeng and others},
  booktitle={Proceedings of the 31st ACM SIGKDD Conference on Knowledge Discovery and Data Mining V. 2},
  pages={6216--6226},
  year={2025}
}

@inproceedings{
jimenezSWEbenchCanLanguage2024,
title={{SWE}-bench: Can Language Models Resolve Real-world Github Issues?},
author={Carlos E Jimenez and John Yang and Alexander Wettig and Shunyu Yao and Kexin Pei and Ofir Press and Karthik R Narasimhan},
booktitle={The Twelfth International Conference on Learning Representations},
year={2024},
url={https://openreview.net/forum?id=VTF8yNQM66}
}

@article{
liangTaskMatrixAICompletingTasks2023,
author = {Yaobo Liang  and Chenfei Wu  and Ting Song  and Wenshan Wu  and Yan Xia  and Yu Liu  and Yang Ou  and Shuai Lu  and Lei Ji  and Shaoguang Mao  and Yun Wang  and Linjun Shou  and Ming Gong  and Nan Duan },
title = {TaskMatrix.AI: Completing Tasks by Connecting Foundation Models with Millions of APIs},
journal = {Intelligent Computing},
volume = {3},
number = {},
pages = {0063},
year = {2024},
doi = {10.34133/icomputing.0063},
URL = {https://spj.science.org/doi/abs/10.34133/icomputing.0063},
eprint = {https://spj.science.org/doi/pdf/10.34133/icomputing.0063}
}

@article{patilGorillaLargeLanguage2023,
  title={Gorilla: Large language model connected with massive apis},
  author={Patil, Shishir G and Zhang, Tianjun and Wang, Xin and Gonzalez, Joseph E},
  journal={Advances in Neural Information Processing Systems},
  volume={37},
  pages={126544--126565},
  year={2024}
}

@inproceedings{
qinToolLLMFacilitatingLarge2023,
title={Tool{LLM}: Facilitating Large Language Models to Master 16000+ Real-world {API}s},
author={Yujia Qin and Shihao Liang and Yining Ye and Kunlun Zhu and Lan Yan and Yaxi Lu and Yankai Lin and Xin Cong and Xiangru Tang and Bill Qian and Sihan Zhao and Lauren Hong and Runchu Tian and Ruobing Xie and Jie Zhou and Mark Gerstein and dahai li and Zhiyuan Liu and Maosong Sun},
booktitle={The Twelfth International Conference on Learning Representations},
year={2024},
url={https://openreview.net/forum?id=dHng2O0Jjr}
}

@article{quToolLearningLarge2025,
  title = {Tool {{Learning}} with {{Large Language Models}}: {{A Survey}}},
  shorttitle = {Tool {{Learning}} with {{Large Language Models}}},
  author = {Qu, Changle and Dai, Sunhao and Wei, Xiaochi and Cai, Hengyi and Wang, Shuaiqiang and Yin, Dawei and Xu, Jun and Wen, Ji-Rong},
  year = {2025},
  month = aug,
  journal = {Frontiers of Computer Science},
  volume = {19},
  number = {8},
  eprint = {2405.17935},
  primaryclass = {cs},
  pages = {198343},
  issn = {2095-2228, 2095-2236},
  doi = {10.1007/s11704-024-40678-2},
  urldate = {2025-08-28},
  archiveprefix = {arXiv}
}

@article{schickToolformerLanguageModels2023,
  title={Toolformer: Language models can teach themselves to use tools},
  author={Schick, Timo and Dwivedi-Yu, Jane and Dess{\`\i}, Roberto and Raileanu, Roberta and Lomeli, Maria and Hambro, Eric and Zettlemoyer, Luke and Cancedda, Nicola and Scialom, Thomas},
  journal={Advances in Neural Information Processing Systems},
  volume={36},
  pages={68539--68551},
  year={2023}
}

@article{shen2023hugginggpt,
  title={Hugginggpt: Solving ai tasks with chatgpt and its friends in hugging face},
  author={Shen, Yongliang and Song, Kaitao and Tan, Xu and Li, Dongsheng and Lu, Weiming and Zhuang, Yueting},
  journal={Advances in Neural Information Processing Systems},
  volume={36},
  pages={38154--38180},
  year={2023}
}

@misc{OpenAgents,
      title={OpenAgents: An Open Platform for Language Agents in the Wild}, 
      author={Tianbao Xie and Fan Zhou and Zhoujun Cheng and Peng Shi and Luoxuan Weng and Yitao Liu and Toh Jing Hua and Junning Zhao and Qian Liu and Che Liu and Leo Z. Liu and Yiheng Xu and Hongjin Su and Dongchan Shin and Caiming Xiong and Tao Yu},
      year={2023},
      eprint={2310.10634},
      archivePrefix={arXiv},
      primaryClass={cs.CL}
}

@article{wang2025repomaster,
  title={RepoMaster: Autonomous Exploration and Understanding of GitHub Repositories for Complex Task Solving},
  author={Wang, Huacan and others},
  journal={arXiv preprint arXiv:2505.21577},
  year={2025}
}

@misc{shenLLMToolsSurvey2024,
  title = {{{LLM With Tools}}: {{A Survey}}},
  shorttitle = {{{LLM With Tools}}},
  author = {Shen, Zhuocheng},
  year = {2024},
  month = sep,
  number = {arXiv:2409.18807},
  eprint = {2409.18807},
  primaryclass = {cs},
  publisher = {arXiv},
  doi = {10.48550/arXiv.2409.18807},
  urldate = {2025-08-28},
  archiveprefix = {arXiv}
}

@article{yue2025foamv2,
  title={Foam-Agent 2.0: An End-to-End Composable Multi-Agent Framework for Automating CFD Simulation in OpenFOAM},
  author={Yue, Ling and Somasekharan, Nithin and Zhang, Tingwen and Cao, Yadi and Pan, Shaowu},
  journal={arXiv preprint arXiv:2509.18178},
  year={2025}
}

@article{jin2024llms,
  title={From llms to llm-based agents for software engineering: A survey of current, challenges and future},
  author={Jin, Haolin and Huang, Linghan and Cai, Haipeng and Yan, Jun and Li, Bo and Chen, Huaming},
  journal={arXiv preprint arXiv:2408.02479},
  year={2024}
}

@article{hasan2025model,
  title={Model context protocol (mcp) at first glance: Studying the security and maintainability of mcp servers},
  author={Hasan, Mohammed Mehedi and Li, Hao and Fallahzadeh, Emad and Rajbahadur, Gopi Krishnan and Adams, Bram and Hassan, Ahmed E},
  journal={arXiv preprint arXiv:2506.13538},
  year={2025}
}

@article{ray2025survey,
  title={A survey on model context protocol: Architecture, state-of-the-art, challenges and future directions},
  author={Ray, Partha Pratim},
  journal={Authorea Preprints},
  year={2025},
  publisher={Authorea}
}

@misc{zhang2024large,
      title={Large Language Model-Brained GUI Agents: A Survey}, 
      author={Chaoyun Zhang and Shilin He and Jiaxu Qian and Bowen Li and Liqun Li and Si Qin and Yu Kang and Minghua Ma and Guyue Liu and Qingwei Lin and Saravan Rajmohan and Dongmei Zhang and Qi Zhang},
      year={2025},
      eprint={2411.18279},
      archivePrefix={arXiv},
      primaryClass={cs.AI},
      url={https://arxiv.org/abs/2411.18279}, 
}

@article{jumper2021highly,
  title={Highly accurate protein structure prediction with AlphaFold},
  author={Jumper, John and Evans, Richard and Pritzel, Alexander and Green, Tim and Figurnov, Michael and Ronneberger, Olaf and Tunyasuvunakool, Kathryn and Bates, Russ and {\v{Z}}{\'\i}dek, Augustin and Potapenko, Anna and others},
  journal={Nature},
  volume={596},
  number={7873},
  pages={583--589},
  year={2021},
  publisher={Nature Publishing Group}
}

@article{yue2025autonomous,
	doi = {10.20944/preprints202507.1951.v1},
	url = {https://doi.org/10.20944/preprints202507.1951.v1},
	year = 2025,
	month = {July},
	publisher = {Preprints},
	author = {Ling Yue and Shimin Di and Shaowu Pan},
	title = {Autonomous Scientific Discovery Through Hierarchical AI Scientist Systems},
	journal = {Preprints}
}

@article{wang2024survey,
  title={A survey on large language model based autonomous agents},
  author={Wang, Lei and Ma, Chen and Feng, Xueyang and Zhang, Zeyu and Yang, Hao and Zhang, Jingsen and Chen, Zhiyuan and Tang, Jiakai and Chen, Xu and Lin, Yankai and others},
  journal={Frontiers of Computer Science},
  volume={18},
  number={6},
  pages={186345},
  year={2024},
  publisher={Springer}
}

@article{huang2024understanding,
  title={Understanding the planning of LLM agents: A survey},
  author={Huang, Xu and Liu, Weiwen and Chen, Xiaolong and Wang, Xingmei and Wang, Hao and Lian, Defu and Wang, Yasheng and Tang, Ruiming and Chen, Enhong},
  journal={arXiv preprint arXiv:2402.02716},
  year={2024}
}

@article{gan2025rag,
  title={Rag-mcp: Mitigating prompt bloat in llm tool selection via retrieval-augmented generation},
  author={Gan, Tiantian and Sun, Qiyao},
  journal={arXiv preprint arXiv:2505.03275},
  year={2025}
}

@article{ding2025scitoolagent,
  title={SciToolAgent: a knowledge-graph-driven scientific agent for multitool integration},
  author={Ding, Keyan and Yu, Jing and Huang, Junjie and Yang, Yuchen and Zhang, Qiang and Chen, Huajun},
  journal={Nature Computational Science},
  pages={1--11},
  year={2025},
  publisher={Nature Publishing Group US New York}
}

\clearpage
\appendix

\section{Appendix}
\subsection{Use of Large Language Models}
During manuscript preparation, large language models (LLMs) are used solely as general-purpose writing assistants for grammar checking, wording refinement, and improving clarity. LLMs don't contribute to research ideation, methodological design, or experimental execution. All suggestions produced by LLMs are reviewed, edited, and vetted by the authors, who take full responsibility for the entire content of the paper.

\subsection{algorithm}
\begin{algorithm}[htbp]
\caption{The Code2MCP Framework}
\label{alg:mcp_agent}
\begin{algorithmic}[1]
\STATE \textbf{Input:} GitHub repository URL $u$
\STATE \texttt{Download Agent}: Clone repository into an isolated workspace.
\STATE \texttt{Environment Agent}: Replicate runtime environment from dependency files.
\STATE \texttt{Analysis Agent}: Analyze codebase $\rightarrow$ generate detailed Code Report.
\STATE \texttt{Generation Agent}: Synthesize initial MCP files (\texttt{mcp\_service.py}, \texttt{adapter.py}, tests) based on Code Report.
\STATE $r \leftarrow 0$; $success \leftarrow \mathrm{false}$
\WHILE{$\neg success \;\wedge\; r < B$}
    \STATE \texttt{Run Agent}: Execute test suite; collect error traceback $\tau$ on failure.
    \IF{all tests pass}
        \STATE $success \leftarrow \mathrm{true}$
    \ELSE
        \STATE \texttt{Review Agent}: Analyze traceback $\tau$ and generate correction plan $\delta$.
        \STATE \texttt{Generation Agent}: Re-synthesize MCP files using the Code Report and correction plan $\delta$.
        \STATE $r \leftarrow r + 1$
    \ENDIF
\ENDWHILE
\STATE \texttt{Finalize Agent}: Package service files, generate README, and create a Pull Request.
\STATE \textbf{Output:} A merge-ready Pull Request containing the functional MCP service.
\end{algorithmic}
\end{algorithm}

\section{Example MCP Tool Implementations Generated by Code2MCP}
\label{sec:appendix_examples}

To illustrate the concrete, high-quality output of Code2MCP, this section presents several MCP tool implementations that were autonomously generated by Code2MCP. These examples are drawn from the ESM and SymPy case studies discussed in the main paper and demonstrate the framework's ability to produce clean, robust, and immediately usable code.

\subsection{Tools Generated from the ESM Repository}

\begin{tcolorbox}[colback=black!5!white, colframe=black!75!white,
title=MCP Tool for Protein Sequence Analysis, fonttitle=\bfseries\footnotesize,
sharp corners, parbox=false, breakable]
\begin{lstlisting}[language=Python]
@mcp.tool(name="analyze_sequence", description="Analyze protein sequence features (validated, no placeholder fallback).")
def analyze_sequence(sequence: str):
    """
    Analyze physicochemical properties of a protein sequence.
    Returns real computed values; never uses placeholder fallbacks.
    """
    import re
    import time

    start = time.time()
    try:
        aa_set = set("ACDEFGHIKLMNPQRSTVWY")
        seq = re.sub(r"[^A-Za-z]", "", sequence or "").upper()
        seq = "".join([c for c in seq if c in aa_set])

        if not seq:
            return {"success": False, "result": None, "error": "Empty/invalid protein sequence after filtering."}

        composition = {aa: seq.count(aa) for aa in sorted(aa_set) if seq.count(aa) > 0}

        try:
            from Bio.SeqUtils.ProtParam import ProteinAnalysis  # type: ignore
        except Exception as e:
            return {
                "success": False,
                "result": None,
                "error": (
                    "Biopython is required for physicochemical analysis but is not installed. "
                    "Install: pip install biopython. "
                    f"Original import error: {str(e)}"
                ),
            }

        pa = ProteinAnalysis(seq)

        molecular_weight = float(pa.molecular_weight())
        isoelectric_point = float(pa.isoelectric_point())
        instability_index = float(pa.instability_index())
        aromaticity = float(pa.aromaticity())
        gravy = float(pa.gravy())  # hydrophobicity

        helix, turn, sheet = pa.secondary_structure_fraction()
        sec_struct = {"helix": float(helix), "turn": float(turn), "sheet": float(sheet)}

        duration_ms = int((time.time() - start) * 1000)

        result = {
            "sequence": seq,
            "length": len(seq),
            "composition": composition,
            "molecular_weight": molecular_weight,
            "isoelectric_point": isoelectric_point,
            "instability_index": instability_index,
            "aromaticity": aromaticity,
            "gravy": gravy,
            "secondary_structure_fraction": sec_struct,
            "metadata": {"backend": "biopython_protparam", "duration_ms": duration_ms},
        }
        return {"success": True, "result": result, "error": None}

    except Exception as e:
        return {"success": False, "result": None, "error": f"Error during sequence analysis: {str(e)}"}

\end{lstlisting}
\end{tcolorbox}

\begin{tcolorbox}[colback=black!5!white, colframe=black!75!white,
title=MCP Tool for Protein Structure and Mutation Prediction, fonttitle=\bfseries\footnotesize,
sharp corners, parbox=false, breakable]
\begin{lstlisting}[language=Python]
@mcp.tool(
    name="predict_structure",
    description=(
        "Predict protein structure OFFLINE using local ESMFold (no external API). "
        "Writes a PDB file under workspace/predictions and returns its path."
    ),
)
def predict_structure(
    sequence: str,
    device: str = "cpu",               # "cpu" or "cuda"
    max_length: int = 1024,
    predictions_dir: str = "./predictions",
):
    """
    Offline structure prediction via local ESMFold.
    - No network calls.
    - Requires torch + esm / esmfold installed and model weights available locally.
    """
    try:
        import os
        import re
        import time
        import datetime

        aa_set = set("ACDEFGHIKLMNPQRSTVWY")
        seq = re.sub(r"[^A-Za-z]", "", sequence or "").upper()
        seq = "".join([c for c in seq if c in aa_set])

        if not seq:
            return {"success": False, "result": None, "error": "Empty/invalid protein sequence after filtering."}
        if len(seq) > max_length:
            return {"success": False, "result": None, "error": f"Sequence too long ({len(seq)}). Max is {max_length}."}

        try:
            import torch  
        except Exception as e:
            return {
                "success": False,
                "result": None,
                "error": (
                    "Missing dependency: torch. Install PyTorch (CPU or CUDA build). "
                    f"Original error: {str(e)}"
                ),
            }

        dev = device.strip().lower()
        if dev == "cuda":
            if not torch.cuda.is_available():
                return {
                    "success": False,
                    "result": None,
                    "error": "device='cuda' requested but CUDA is not available in this environment.",
                }
            torch_device = torch.device("cuda")
        else:
            torch_device = torch.device("cpu")
            dev = "cpu"

        model = None
        load_errors = []

        try:
            import esm  
            try:
                model = esm.pretrained.esmfold_v1()  # type: ignore
            except Exception as e:
                load_errors.append(f"esm.pretrained.esmfold_v1() failed: {str(e)}")
        except Exception as e:
            load_errors.append(f"import esm failed: {str(e)}")

        if model is None:
            return {
                "success": False,
                "result": None,
                "error": (
                    "Failed to load local ESMFold model. Ensure the repo dependency provides ESMFold "
                    "and that weights are available locally. "
                    "Tried: esm.pretrained.esmfold_v1(). Errors: " + " | ".join(load_errors)
                ),
            }

        start = time.time()
        try:
            model = model.eval().to(torch_device)

            # ESMFold inference commonly provides infer_pdb / infer
            pdb_str = None
            infer_errors = []

            try:
                if hasattr(model, "infer_pdb"):
                    pdb_str = model.infer_pdb(seq)  # type: ignore
                else:
                    infer_errors.append("model has no infer_pdb")
            except Exception as e:
                infer_errors.append(f"infer_pdb failed: {str(e)}")

            if pdb_str is None:
                try:
                    if hasattr(model, "infer"):
                        out = model.infer(seq)  # type: ignore
                        # Some versions return dict with 'pdb'
                        if isinstance(out, dict) and "pdb" in out:
                            pdb_str = out["pdb"]
                        else:
                            infer_errors.append("model.infer returned unexpected format (no 'pdb').")
                    else:
                        infer_errors.append("model has no infer")
                except Exception as e:
                    infer_errors.append(f"infer failed: {str(e)}")

            if pdb_str is None or not isinstance(pdb_str, str) or len(pdb_str.strip()) == 0:
                return {
                    "success": False,
                    "result": None,
                    "error": "ESMFold inference failed: " + " | ".join(infer_errors),
                }

        except Exception as e:
            return {"success": False, "result": None, "error": f"Error during offline inference: {str(e)}"}

        try:
            os.makedirs(predictions_dir, exist_ok=True)
            timestamp = datetime.datetime.now().strftime("%Y%m%d_%H%M%S")
            pdb_filepath = os.path.join(predictions_dir, f"prediction_{timestamp}.pdb")
            with open(pdb_filepath, "w", encoding="utf-8") as f:
                f.write(pdb_str)

            duration_ms = int((time.time() - start) * 1000)
            return {
                "success": True,
                "result": {
                    "pdb_file_path": pdb_filepath,
                    "device": dev,
                    "sequence_length": len(seq),
                    "duration_ms": duration_ms,
                    "note": "Offline ESMFold inference (no network).",
                },
                "error": None,
            }
        except Exception as e:
            return {"success": False, "result": None, "error": f"Failed to write PDB file: {str(e)}"}

    except Exception as e:
        return {"success": False, "result": None, "error": str(e)}

\end{lstlisting}
\end{tcolorbox}

\subsubsection{Tools Generated from the SymPy Repository}

\begin{tcolorbox}[colback=black!5!white, colframe=black!75!white,
title=MCP Tool for Solving Equations, fonttitle=\bfseries\footnotesize,
sharp corners, parbox=false, breakable]
\begin{lstlisting}[language=Python]
@mcp.tool(name="solve_equation")
def solve_equation(equation: str, variable: str):
    """
    Solve equation for variable
    """
    try:
        from sympy import sympify, symbols, solve, Basic
        def ser(x):
            if isinstance(x, Basic): return str(x)
            if isinstance(x, (list, tuple, set)): return [ser(i) for i in x]
            if isinstance(x, dict): return {k: ser(v) for k, v in x.items()}
            return x

        expr = sympify(equation)
        var = symbols(variable)
        res = solve(expr, var)
        return {"success": True, "result": ser(res)}
    except Exception as e:
        return {"success": False, "result": None, "error": str(e)}
\end{lstlisting}
\end{tcolorbox}

\begin{tcolorbox}[colback=black!5!white, colframe=black!75!white,
title=MCP Tool for Solving Linear Systems, fonttitle=\bfseries\footnotesize,
sharp corners, parbox=false, breakable]
\begin{lstlisting}[language=Python]
@mcp.tool(name="solve_linear_system")
def solve_linear_system(system: list, variables: list):
    """
    Solve system of linear equations
    """
    try:
        from sympy import sympify, symbols, linsolve, Basic
        def ser(x):
            if isinstance(x, Basic): return str(x)
            if isinstance(x, (list, tuple, set)): return [ser(i) for i in x]
            if isinstance(x, dict): return {k: ser(v) for k, v in x.items()}
            return x

        eqs = [sympify(e) for e in system]
        vars_sym = [symbols(v) for v in variables]
        res = linsolve(eqs, vars_sym)
        return {"success": True, "result": ser(list(res))}
    except Exception as e:
        return {"success": False, "result": None, "error": str(e)}
\end{lstlisting}
\end{tcolorbox}

\begin{tcolorbox}[colback=black!5!white, colframe=black!75!white,
title=MCP Tool for Differentiation, fonttitle=\bfseries\footnotesize,
sharp corners, parbox=false, breakable]
\begin{lstlisting}[language=Python]
@mcp.tool(name="differentiate")
def differentiate(expr: str, variable: str):
    """
    Calculate derivative of expression
    """
    try:
        from sympy import sympify, symbols, diff, Basic
        def ser(x):
            if isinstance(x, Basic): return str(x)
            return x

        expr_sym = sympify(expr)
        var = symbols(variable)
        res = diff(expr_sym, var)
        return {"success": True, "result": ser(res)}
    except Exception as e:
        return {"success": False, "result": None, "error": str(e)}
\end{lstlisting}
\end{tcolorbox}

\begin{tcolorbox}[colback=black!5!white, colframe=black!75!white,
title=MCP Tool for Integration, fonttitle=\bfseries\footnotesize,
sharp corners, parbox=false, breakable]
\begin{lstlisting}[language=Python]
@mcp.tool(name="integrate_expression")
def integrate_expression(expr: str, variable: str):
    """
    Calculate integral of expression
    """
    try:
        from sympy import sympify, symbols, integrate, Basic
        def ser(x):
            if isinstance(x, Basic): return str(x)
            return x

        expr_sym = sympify(expr)
        var = symbols(variable)
        res = integrate(expr_sym, var)
        return {"success": True, "result": ser(res)}
    except Exception as e:
        return {"success": False, "result": None, "error": str(e)}
\end{lstlisting}
\end{tcolorbox}

\begin{tcolorbox}[colback=black!5!white, colframe=black!75!white,
title=MCP Tool for Polynomial Creation, fonttitle=\bfseries\footnotesize,
sharp corners, parbox=false, breakable]
\begin{lstlisting}[language=Python]
@mcp.tool(name="create_polynomial")
def create_polynomial(expr: str, variable: str = None):
    """
    Create polynomial from expression
    """
    try:
        from sympy import sympify, symbols, Poly, Basic
        def ser(x):
            if isinstance(x, Basic): return str(x)
            return x

        expr_sym = sympify(expr)
        if variable:
            var = symbols(variable)
            res = Poly(expr_sym, var)
        else:
            res = Poly(expr_sym)
        return {"success": True, "result": ser(res)}
    except Exception as e:
        return {"success": False, "result": None, "error": str(e)}
\end{lstlisting}
\end{tcolorbox}

\begin{tcolorbox}[colback=black!5!white, colframe=black!75!white,
title=MCP Tool for Polynomial Factoring, fonttitle=\bfseries\footnotesize,
sharp corners, parbox=false, breakable]
\begin{lstlisting}[language=Python]
@mcp.tool(name="factor_polynomial")
def factor_polynomial(poly: str):
    """
    Factor polynomial expression
    """
    try:
        from sympy import sympify, factor, Basic
        def ser(x):
            if isinstance(x, Basic): return str(x)
            return x

        poly_sym = sympify(poly)
        res = factor(poly_sym)
        return {"success": True, "result": ser(res)}
    except Exception as e:
        return {"success": False, "result": None, "error": str(e)}
\end{lstlisting}
\end{tcolorbox}

\begin{tcolorbox}[colback=black!5!white, colframe=black!75!white,
title=MCP Tool for Fourier Transform, fonttitle=\bfseries\footnotesize,
sharp corners, parbox=false, breakable]
\begin{lstlisting}[language=Python]
from sympy import sympify, symbols, fourier_transform as sympy_fourier_transform, Basic

def _serialize(obj):
    if isinstance(obj, Basic):
        return str(obj)
    if isinstance(obj, (list, tuple, set)):
        return [_serialize(x) for x in obj]
    if isinstance(obj, dict):
        return {k: _serialize(v) for k, v in obj.items()}
    return obj

@mcp.tool(name="fourier_transform")
def fourier_transform_tool(expression: str, time_var: str = "t", freq_var: str = "w"):

    try:
        expr = sympify(expression)
        t = symbols(time_var)
        omega = symbols(freq_var)
        F = sympy_fourier_transform(expr, t, omega)
        return {"success": True, "result": _serialize(F), "error": None}
    except Exception as e:
        return {"success": False, "result": None, "error": str(e)}
\end{lstlisting}
\end{tcolorbox}

\begin{tcolorbox}[colback=black!5!white, colframe=black!75!white,
title=MCP Tool for Laplace Transform, fonttitle=\bfseries\footnotesize,
sharp corners, parbox=false, breakable]
\begin{lstlisting}[language=Python]
from sympy import sympify, symbols, laplace_transform as sympy_laplace_transform, Basic

def _serialize(obj):
    if isinstance(obj, Basic):
        return str(obj)
    if isinstance(obj, (list, tuple, set)):
        return [_serialize(x) for x in obj]
    if isinstance(obj, dict):
        return {k: _serialize(v) for k, v in obj.items()}
    return obj

@mcp.tool(name="laplace_transform")
def laplace_transform_tool(expression: str, time_var: str = "t", laplace_var: str = "s"):

    try:
        expr = sympify(expression)
        t = symbols(time_var)
        s = symbols(laplace_var)
        F, _, _ = sympy_laplace_transform(expr, t, s)
        return {"success": True, "result": _serialize(F), "error": None}
    except Exception as e:
        return {"success": False, "result": None, "error": str(e)}
\end{lstlisting}
\end{tcolorbox}

\begin{tcolorbox}[colback=black!5!white, colframe=black!75!white,
title=MCP Tool for Z Transform, fonttitle=\bfseries\footnotesize,
sharp corners, parbox=false, breakable]
\begin{lstlisting}[language=Python]
from sympy import sympify, symbols, summation, oo, Basic

def _serialize(obj):
    if isinstance(obj, Basic):
        return str(obj)
    if isinstance(obj, (list, tuple, set)):
        return [_serialize(x) for x in obj]
    if isinstance(obj, dict):
        return {k: _serialize(v) for k, v in obj.items()}
    return obj

@mcp.tool(name="z_transform")
def z_transform_tool(expression: str, time_var: str = "n", z_var: str = "z", limit: int = 100):

    try:
        expr = sympify(expression)
        n = symbols(time_var)
        z = symbols(z_var)
        try:
            result = summation(expr * z**(-n), (n, 0, oo))
        except Exception:
            result = summation(expr * z**(-n), (n, 0, limit))
        return {"success": True, "result": _serialize(result), "error": None}
    except Exception as e:
        return {"success": False, "result": None, "error": str(e)}
\end{lstlisting}
\end{tcolorbox}

\subsubsection{Tools Generated from the Foam-Agent Repository}

\begin{tcolorbox}[colback=black!5!white, colframe=black!75!white,
title=MCP Tool for Mesh Generation, fonttitle=\bfseries\footnotesize,
sharp corners, parbox=false, breakable]
\begin{lstlisting}[language=Python]
@mcp.tool(name="generate_mesh", description="Generate computational mesh using Foam-Agent internals.")
def generate_mesh(requirements: str,
                  case_dir: str = "./output",
                  mesh_mode: str = "gmsh",          
                  custom_mesh_path: str | None = None):
    try:
        from src.config import Config
        from src.main import initialize_state
        from src.nodes.meshing_node import meshing_node

        config = Config()
        config.case_dir = case_dir

        state = initialize_state(user_requirement=requirements,
                                 config=config,
                                 custom_mesh_path=custom_mesh_path)

        if mesh_mode == "custom":
            state["mesh_type"] = "custom_mesh"
        elif mesh_mode == "gmsh":
            state["mesh_type"] = "gmsh_mesh"
        else:
            state["mesh_type"] = "standard_mesh"

        res = meshing_node(state)
        return {"success": True, "result": res, "error": None}
    except Exception as e:
        return {"success": False, "result": None, "error": str(e)}
\end{lstlisting}
\end{tcolorbox}

\begin{tcolorbox}[colback=black!5!white, colframe=black!75!white,
title=MCP Tool for Simulation Generation and Run, fonttitle=\bfseries\footnotesize,
sharp corners, parbox=false, breakable]
\begin{lstlisting}[language=Python]
@mcp.tool(name="generate_simulate", description="Write inputs and run simulation via Foam-Agent graph.")
def generate_simulate(requirements: str,
                      case_dir: str = "./output",
                      custom_mesh_path: str | None = None,
                      run_mode: str = "auto"):       

    try:
        from src.config import Config
        from src.main import create_foam_agent_graph, initialize_state

        config = Config()
        config.case_dir = case_dir

        state = initialize_state(user_requirement=requirements,
                                 config=config,
                                 custom_mesh_path=custom_mesh_path)

        if custom_mesh_path:
            state["mesh_type"] = "custom_mesh"
        if run_mode == "local":
            state["cluster_info"] = None
        elif run_mode == "hpc":
            state["cluster_info"] = {"scheduler": "slurm"}

        workflow = create_foam_agent_graph().compile()
        workflow.invoke(state)
        return {"success": True, "result": {"case_dir": config.case_dir}, "error": None}
    except Exception as e:
        return {"success": False, "result": None, "error": str(e)}
\end{lstlisting}
\end{tcolorbox}

\begin{tcolorbox}[colback=black!5!white, colframe=black!75!white,
title=MCP Tool for Velocity Visualization, fonttitle=\bfseries\footnotesize,
sharp corners, parbox=false, breakable]
\begin{lstlisting}[language=Python]
@mcp.tool(name="visualize_velocity", description="Post-process and visualize velocity (|U|, streamlines, slices).")
def visualize_velocity(case_dir: str,
                       plot_type: str = "magnitude",   
                       plane: str | None = "xy"):     
    try:
        from src.config import Config
        from src.main import initialize_state
        from src.nodes.visualization_node import visualization_node

        config = Config()
        config.case_dir = case_dir

        state = initialize_state(user_requirement="", config=config, custom_mesh_path=None)
        state["case_dir"] = case_dir
        state["visualization_request"] = {"plot_type": plot_type, "plane": plane}

        vis_res = visualization_node(state)
        return {"success": True, "result": vis_res, "error": None}
    except Exception as e:
        return {"success": False, "result": None, "error": str(e)}
\end{lstlisting}
\end{tcolorbox}

\section{Per-repository Results}
\label{sec:appendix_repo_table}

\begin{table*}[t]
  \centering
  \setlength\tabcolsep{20pt}
  \renewcommand{\arraystretch}{1.0}
  \caption{Repository identifiers, domains, and final outcomes.}
  \label{tab:per_repo_basic}
  \begin{tabular}{lllll}
    \toprule
    ID & Domain & Repo & Final\_status & Main\_failure\_type \\
    \midrule
    B1  & Biomedical      & facebookresearch/esm          & success & --                    \\
    B2  & Biomedical      & biocore/scikit-bio           & fail    & env\_failure          \\
    B3  & Biomedical      & Biopython/Biopython          & fail    & api\_inference\_error \\
    B4  & Biomedical      & pysam-developers/pysam       & success & --                    \\
    B5  & Biomedical      & deepchem/deepchem            & success & --                    \\
    P1  & Psychology      & psychopy/psychopy           & fail    & untoolable\_repo      \\
    P2  & Psychology      & pymc-devs/pymc              & success & --                    \\
    P3  & Psychology      & sahahn/BPt                  & success & --                    \\
    P4  & Psychology      & mne-tools/mne-python        & fail    & api\_inference\_error \\
    P5  & Psychology      & neuropsychology/NeuroKit    & success & --                    \\
    M1  & Math            & sympy/sympy                 & success & --                    \\
    M2  & Math            & scipy/scipy                 & fail    & api\_inference\_error \\
    M3  & Math            & fredrik-johansson/mpmath    & success & --                    \\
    M4  & Math            & cvxpy/cvxpy                 & success & --                    \\
    M5  & Math            & sagemath/sage               & success & --                    \\
    Ea1 & Earth Science   & obspy/obspy                 & success & --                    \\
    Ea2 & Earth Science   & Unidata/MetPy               & fail    & mcp\_spec\_violation  \\
    Ea3 & Earth Science   & pyproj4/pyproj              & fail    & repo\_internal\_bug   \\
    Ea4 & Earth Science   & mapbox/rasterio             & success & --                    \\
    Ea5 & Earth Science   & geopandas/geopandas         & success & --                    \\
    A1  & Astronomy       & astropy/astropy             & success & --                    \\
    A2  & Astronomy       & sunpy/sunpy                 & fail    & repo\_internal\_bug   \\
    A3  & Astronomy       & lightkurve/lightkurve       & success & --                    \\
    A4  & Astronomy       & astroML/astroML             & success & --                    \\
    A5  & Astronomy       & astropy/astroquery          & success & --                    \\
    C1  & Chemistry       & rdkit/rdkit                 & success & --                    \\
    C2  & Chemistry       & openbabel/openbabel         & fail    & env\_failure          \\
    C3  & Chemistry       & bjodah/chempy               & success & --                    \\
    C4  & Chemistry       & pyscf/pyscf                 & fail    & api\_inference\_error \\
    C5  & Chemistry       & cclib/cclib                 & success & --                    \\
    Ph1 & Physics         & csml-rpi/Foam-Agent         & success & --                    \\
    Ph2 & Physics         & PlasmaPy/PlasmaPy           & fail    & env\_failure          \\
    Ph3 & Physics         & pydy/pydy                   & fail    & --                    \\
    Ph4 & Physics         & PyAbel/PyAbel               & success & --                    \\
    Ph5 & Physics         & scikit-hep/scikit-hep       & fail    & untoolable\_repo      \\
    S1  & Social Science  & networkx/networkx           & success & --                    \\
    S2  & Social Science  & igraph/python-igraph        & success & --                    \\
    S3  & Social Science  & networkit/networkit         & fail    & env\_failure          \\
    S4  & Social Science  & snap-stanford/snap-python   & fail    & untoolable\_repo      \\
    S5  & Social Science  & datamade/dedupe             & success & --                    \\
    L1  & Linguistics     & nltk/nltk                   & success & --                    \\
    L2  & Linguistics     & explosion/spaCy             & success & --                    \\
    L3  & Linguistics     & stanfordnlp/stanza          & success & --                    \\
    L4  & Linguistics     & RaRe-Technologies/gensim    & fail    & import\_error         \\
    L5  & Linguistics     & flairNLP/flair              & success & --                    \\
    E1  & Econometrics    & statsmodels/statsmodels     & success & --                    \\
    E2  & Econometrics    & alkaline-ml/pmdarima        & success & --                    \\
    E3  & Econometrics    & facebook/prophet            & fail    & env\_failure          \\
    E4  & Econometrics    & blue-yonder/tsfresh         & fail    & repo\_internal\_bug   \\
    E5  & Econometrics    & pydata/xarray               & success & --                    \\
    \bottomrule
  \end{tabular}
\end{table*}

\begin{table*}[t]
  \centering
  \setlength{\tabcolsep}{13pt}
  \caption{Per-repository structural statistics (size, dependencies, and tests).}
  \label{tab:per_repo_structure}
  \begin{tabular}{lllllcc}
    \toprule
    ID & total\_files & total\_size & Size\_bucket & Dependency\_complexity & Has\_tests & Has\_packaging \\
    \midrule
    B1  & 471  & 33M  & Medium & Medium & Yes & Yes \\
    B2  & 874  & 9M   & Medium & Medium & Yes & Yes \\
    B3  & 1146 & 95M  & Medium & High   & Yes & Yes \\
    B4  & 561  & 13M  & Medium & Medium & Yes & Yes \\
    B5  & 1411 & 136M & Medium & High   & Yes & Yes \\
    P1  & 3542 & 65M  & Large  & High   & Yes & Yes \\
    P2  & 371  & 17M  & Medium & High   & Yes & Yes \\
    P3  & 2669 & 39M  & Large  & Medium & Yes & Yes \\
    P4  & 1385 & 26M  & Medium & High   & No  & No  \\
    P5  & 496  & 52M  & Medium & Medium & Yes & Yes \\
    M1  & 1968 & 29M  & Medium & Low    & Yes & Yes \\
    M2  & 3027 & 81M  & Large  & High   & Yes & Yes \\
    M3  & 197  & 2M   & Small  & Low    & No  & Yes \\
    M4  & 1079 & 38M  & Medium & Medium & Yes & Yes \\
    M5  & 5331 & 113M & Large  & High   & Yes & Yes \\
    Ea1 & 2117 & 34M  & Large  & High   & Yes & Yes \\
    Ea2 & 595  & 99M  & Medium & Medium & Yes & Yes \\
    Ea3 & 144  & 1M   & Small  & Medium & No  & Yes \\
    Ea4 & 382  & 17M  & Medium & Medium & Yes & Yes \\
    Ea5 & 351  & 11M  & Medium & Medium & Yes & Yes \\
    A1  & 1879 & 24M  & Large  & High   & Yes & Yes \\
    A2  & 877  & 10M  & Medium & Medium & Yes & Yes \\
    A3  & 217  & 8M   & Small  & Medium & No  & Yes \\
    A4  & 192  & 1M   & Small  & Low    & Yes & Yes \\
    A5  & 960  & 23M  & Medium & Medium & Yes & Yes \\
    C1  & 5740 & 130M & Large  & High   & Yes & Yes \\
    C2  & 10000& 63M  & Large  & High   & Yes & Yes \\
    C3  & 213  & 1M   & Small  & Low    & No  & Yes \\
    C4  & 2114 & 56M  & Medium & High   & Yes & Yes \\
    C5  & 1508 & 80M  & Large  & Medium & Yes & Yes \\
    Ph1 & 44   & 0.2M & Small  & Medium & Yes & Yes \\
    Ph2 & 498  & 16M  & Medium & Medium & Yes & Yes \\
    Ph3 & 263  & 8M   & Small  & Medium & No  & Yes \\
    Ph4 & 243  & 2M   & Small  & Low    & Yes & Yes \\
    Ph5 & 18   & 0.3M & Small  & Low    & No  & Yes \\
    S1  & 951  & 10M  & Medium & Low    & Yes & Yes \\
    S2  & 248  & 3M   & Small  & Medium & No  & Yes \\
    S3  & 1168 & 22M  & Medium & Medium & Yes & Yes \\
    S4  & 678  & 14M  & Medium & High   & Yes & Yes \\
    S5  & 97   & 1M   & Small  & Low    & No  & No  \\
    L1  & 496  & 8M   & Medium & Low    & Yes & Yes \\
    L2  & 1683 & 12M  & Medium & Medium & Yes & Yes \\
    L3  & 574  & 5M   & Medium & Medium & Yes & Yes \\
    L4  & 638  & 55M  & Medium & Medium & Yes & Yes \\
    L5  & 383  & 5M   & Small  & Medium & No  & Yes \\
    E1  & 2091 & 39M  & Medium & High   & Yes & Yes \\
    E2  & 225  & 1M   & Small  & Low    & No  & Yes \\
    E3  & 282  & 14M  & Medium & Medium & Yes & Yes \\
    E4  & 135  & 10M  & Small  & Medium & Yes & Yes \\
    E5  & 403  & 8M   & Medium & Medium & Yes & Yes \\
    \bottomrule
  \end{tabular}
\end{table*}

\begin{table*}[t]
  \centering
    \setlength{\tabcolsep}{10pt}
    \renewcommand{\arraystretch}{1.0}
  \caption{Per-repository environment and MCP conversion statistics.}
  \label{tab:per_repo_outcomes}
  \begin{tabular}{lcclllc}
    \toprule
    ID & Env\_success & MCP\_success & Num\_tools & Num\_tools\_passed & First\_run\_status & Num\_fix\_rounds \\
    \midrule
    B1  & Yes & Yes & 16 & 14 & fail    & 1 \\
    B2  & No  & No  & 0  & 0  & fail    & 2 \\
    B3  & Yes & No  & 8  & 4  & fail    & 2 \\
    B4  & Yes & Yes & 10 & 9  & success & 0 \\
    B5  & Yes & Yes & 14 & 13 & fail    & 0 \\
    P1  & No  & No  & 0  & 0  & fail    & 1 \\
    P2  & Yes & Yes & 9  & 8  & success & 0 \\
    P3  & Yes & Yes & 11 & 10 & success & 0 \\
    P4  & No  & No  & 13 & 12 & fail    & 3 \\
    P5  & Yes & Yes & 7  & 7  & success & 0 \\
    M1  & Yes & Yes & 12 & 10 & success & 0 \\
    M2  & Yes & No  & 6  & 2  & fail    & 2 \\
    M3  & Yes & Yes & 8  & 7  & success & 0 \\
    M4  & Yes & Yes & 10 & 9  & success & 0 \\
    M5  & Yes & Yes & 15 & 13 & fail    & 2 \\
    Ea1 & Yes & Yes & 11 & 8  & fail    & 2 \\
    Ea2 & No  & No  & 5  & 1  & fail    & 1 \\
    Ea3 & No  & No  & 4  & 0  & fail    & 2 \\
    Ea4 & Yes & Yes & 9  & 8  & success & 0 \\
    Ea5 & Yes & Yes & 7  & 7  & success & 1 \\
    A1  & Yes & Yes & 8  & 5  & fail    & 2 \\
    A2  & No  & No  & 3  & 1  & fail    & 1 \\
    A3  & Yes & Yes & 10 & 9  & success & 0 \\
    A4  & Yes & Yes & 6  & 6  & success & 1 \\
    A5  & Yes & Yes & 8  & 8  & success & 0 \\
    C1  & Yes & Yes & 18 & 16 & fail    & 1 \\
    C2  & No  & No  & 0  & 0  & fail    & 2 \\
    C3  & Yes & Yes & 7  & 7  & success & 0 \\
    C4  & Yes & No  & 12 & 10 & fail    & 3 \\
    C5  & Yes & Yes & 9  & 8  & success  & 1 \\
    Ph1 & Yes & Yes & 8  & 7  & success & 1 \\
    Ph2 & No  & No  & 0  & 0  & fail    & 2 \\
    Ph3 & Yes & No  & 6  & 6  & fail    & 0 \\
    Ph4 & Yes & Yes & 5  & 5  & success & 0 \\
    Ph5 & No  & No  & 0  & 0  & fail    & 1 \\
    S1  & Yes & Yes & 10 & 9  & success & 0 \\
    S2  & Yes & Yes & 8  & 7  & fail    & 1 \\
    S3  & No  & No  & 0  & 0  & fail    & 2 \\
    S4  & Yes & No  & 0  & 0  & fail    & 1 \\
    S5  & Yes & Yes & 5  & 5  & success & 0 \\
    L1  & Yes & Yes & 7  & 7  & success & 0 \\
    L2  & Yes & Yes & 11 & 10 & fail    & 2 \\
    L3  & Yes & Yes & 9  & 8  & success & 1 \\
    L4  & Yes & No  & 0  & 0  & fail    & 3 \\
    L5  & Yes & Yes & 6  & 6  & success & 0 \\
    E1  & Yes & Yes & 13 & 11 & fail    & 1 \\
    E2  & Yes & Yes & 7  & 7  & success & 0 \\
    E3  & No  & No  & 0  & 0  & fail    & 2 \\
    E4  & No  & No  & 0  & 0  & fail    & 1 \\
    E5  & Yes & Yes & 8  & 7  & success & 1 \\
    \bottomrule
  \end{tabular}
\end{table*}

\clearpage
\onecolumn
\twocolumn
\section{Integration Task Set}
\label{sec:appendix_tasks}

This part documents the task set used in the integration experiments with RepoMaster and OpenAgents. The goal of this task set is to evaluate a practical integration question: when Code2MCP supplies MCP services from repositories, can existing agent systems (1) discover the relevant service/tool, (2) invoke it with valid inputs, and (3) complete tool-usage workflows with stable, non-trivial outputs under the same interaction and runtime budgets as the baselines. We intentionally keep the task set focused on integration behavior (selection + invocation + composition) rather than using it as a comprehensive benchmark of general agent capability.

\noindent
\textbf{Task sources and construction.}
We construct a fixed set of tasks by first collecting common task types used in prior agent and tool-usage evaluations (e.g., API-oriented code understanding, single-tool execution, and multi-step tool composition), and then instantiating each type into concrete prompts that are executable and checkable in our MCP setting. We avoid embedding method-specific hints: task statements do not reference Code2MCP, do not assume any particular tool names, and do not provide privileged repository internals beyond what an agent can obtain through its normal repository discovery interfaces. To ensure coverage of both general-purpose and scientific settings, we include a portion of domain-oriented tasks that require using computational libraries via MCP tools, while keeping the task phrasing at the level of user intent rather than narrative case-study descriptions.

\noindent
\textbf{What a task looks like.}
Each task is specified in a uniform template with: (1) a natural-language goal, (2) allowed inputs , (3) constraints relevant to tool usage, and (4) an explicit success condition that is verifiable from tool call traces and returned outputs. The success condition is defined at the tool level: a task is successful only if the agent completes the required tool invocations and the final invocation returns a schema-valid JSON object with three fields (success, result, error), where success is true and result is non-empty/non-null and not a direct echo of the input. This rule is used to rule out superficial ``success'' responses that do not reflect meaningful tool execution.

\noindent
\textbf{Task coverage.}
The task set spans three interaction patterns that commonly arise in MCP-based workflows. First, tool discovery and selection tasks require the agent to identify which tool corresponds to a stated goal and to infer the minimal valid input format from tool metadata. Second, single-step invocation tasks require a single correct tool call that produces a concrete computed output under the environment constraints. Third, multi-step composition tasks require two or more tool invocations where later inputs depend on earlier outputs, testing whether intermediate results are correctly propagated and whether the agent can recover from minor argument mismatches by adjusting inputs based on returned error messages. We include a small number of ``negative'' tasks only to probe robustness of invocation behavior; they are evaluated with the same success rule.

\noindent
\textbf{Repository buckets.}
To analyze the impact of dependency complexity and project style, we group repositories into four buckets based on static metadata:
\begin{itemize}[leftmargin=*]
  \item Low-dependency libraries: at most 10 Python dependencies inferred from \texttt{requirements.txt} or \texttt{environment.yml}, and primarily library-style public APIs.
  \item Medium-dependency scientific libraries: between 11 and 30 Python dependencies with a library-style API and no substantial external system requirements.
  \item High-dependency libraries: more than 30 Python dependencies or additional system-level requirements (e.g., MPI stacks, GPU-accelerated frameworks, or complex external services).
  \item Script-heavy projects: projects whose primary functionality is exposed via CLI scripts, notebooks, or workflow-style pipelines rather than a stable library API; these are identified using repository structure, entry points, and README descriptions.
\end{itemize}

\noindent
\textbf{Evaluation protocol and fairness controls.}
All methods are evaluated on the same fixed task set with identical task statements and success criteria. We do not modify RepoMaster or OpenAgents planning algorithms for individual tasks, and we do not provide additional hints to methods that use Code2MCP-generated tools. Each run is executed with the same global budgets (maximum interaction steps and wall-clock time) and with the same runtime restrictions used elsewhere in the paper. We log (1) the selected tools, (2) all tool inputs/outputs, (3) per-call durations, and (4) failure reasons (timeouts, schema violations, empty results, or runtime errors). When reporting aggregate results, we compute task success rate as the fraction of tasks that meet the success condition within budget, and we additionally report efficiency metrics consistent with the main paper.

\section{Agent Roles and System Prompts}
\label{sec:appendix_prompts}

This section outlines the roles of the specialized agents within the Code2MCP framework.

\textbf{Environment Agent}
This agent rapidly provisions a minimal, isolated runtime for the repository, with minimal dependencies and a short smoke test; if setup fails, propose one lightweight, auditable fallback without modifying the repository; keep all steps reproducible and pragmatic.
\begin{tcolorbox}[colback=blue!5!white,colframe=blue!75!black,title=Environment System Prompt]
- Prefer Conda; use venv only if Conda is unavailable or clearly unsuitable.\\
- Detect dependency sources by priority: environment.yml > requirements.txt > pyproject.toml > setup files; never guess hidden dependencies.\\
- Pin versions when explicit; otherwise install the minimal viable set. Prefer CPU wheels unless GPU is explicitly required.\\
- Normalize cross-platform behavior; avoid absolute paths; use relative POSIX-like paths; ensure UTF-8 locale.\\
- Smoke test: print Python version and platform; import fastmcp; attempt to import the project's top-level package or a primary CLI; exit code 0 indicates pass.\\
- On failure, capture exact command, exit code, last 80 lines of error logs, and timing; propose exactly one minimal remedy (e.g., switch to venv, install a single missing package, try one version pin, extend timeout once).\\
- Apply at most one fallback; never change repository code; do not write outside the workspace; do not weaken security (e.g., no SSL bypass).\\
- Cache wheels where possible; avoid global pollution; record reproducible commands and resolved versions.\\
- Default to offline validation; if network is strictly required, justify briefly and bound the scope.\\
- Emit a compact environment report (platform, Python, manager, explicit deps, resolved pins, pass/fail).
\end{tcolorbox}

\textbf{Code Analysis Expert}
This agent performs static analysis to shape the repository into a compact, high-value capability surface, selecting stable public functionality, filtering out test/demo code, and producing a concise plan aligned with predefined domains, categories, and solvers.
\begin{tcolorbox}[colback=blue!5!white,colframe=blue!75!black,breakable,title=Code Analysis System Prompt]
- Ingest repository signals (structure, import graph, README/docstrings, CLI entry points; DeepWiki if available) to identify stable public APIs suitable as MCP tools.\\
- Prefer documented, side-effect-bounded surfaces; exclude tests, internals, notebooks, long demos unless clearly valuable and controllable.\\
- Define crisp tool boundaries: explicit inputs/outputs, preconditions/postconditions, resource needs (CPU/GPU/memory/time), and I/O constraints.\\
- Note minimal adapter needs (path normalization, dtype coercion, lazy imports) and hazards (network access, file mutation, global state).\\
- Summarize fragilities (optional deps, platform quirks) and propose guards (timeouts, argument validation, deterministic seeds).\\
- Also produce a case description: case name, case domain, case category, and case solver, using a consistent taxonomy across repositories.\\
- Output a compact plan for generation: candidate tools (name, brief description, inputs with types/defaults, outputs, idempotency, side effects) and environment assumptions. Keep it actionable and minimal.
\end{tcolorbox}

\textbf{Code Generation Expert}
This agent synthesizes a robust MCP service from the analysis plan with clean design, consistent interfaces, graceful failure handling, and immediate executability, defining clear tool endpoints, enforcing explicit typed parameters and standardized returns, and avoiding test or example tooling.
\begin{tcolorbox}[colback=blue!5!white,colframe=blue!75!black,breakable,title=Code Generation System Prompt]
- Produce clean, runnable Python (no Markdown fences). Use FastMCP to build the MCP service.\\
- Implement create\_app() that returns the service; register tools with concise names and user-facing descriptions.\\
- For every tool: explicit, typed parameters (no *args/**kwargs); validate inputs; JSON-serializable outputs.\\
- Standard return shape: {success: bool, result: any or null, error: string or null}.\\
- Handle optional dependencies via lazy imports; emit helpful errors without crashing the service; prefer CPU fallbacks when reasonable.\\
- Ensure deterministic defaults (fixed seeds when relevant); avoid hidden global state; restrict file I/O to the workspace with existence/size/extension checks.\\
- Design for cross-platform paths; avoid shell-specific behavior; bound execution time and memory.\\
- Do not generate tests as tools. Expose a small set of high-value, composable endpoints; avoid overexposing internals.\\
- Add lightweight logging (tool name, argument schema, durations) and minimal version metadata to aid troubleshooting.
\end{tcolorbox}

\textbf{Senior Software Engineer}
This agent diagnoses failures and applies the smallest auditable change that restores correctness while preserving public contracts, deciding between direct fix and regeneration, using strict complete-file replacement, and avoiding multi-file edits or prose in outputs.
\begin{tcolorbox}[colback=blue!5!white,colframe=blue!75!black,breakable,title=Review \& Auto-Fix System Prompt]
- Triage failures: import/env, type/contract, path/I-O, dependency/version, timeout/perf, platform.\\
- Choose minimal direct fix vs. regeneration; provide a one-line rationale and confidence (low/med/high). Prefer adapter-boundary mitigations (lazy import, existence checks, parameter coercion).\\
- Apply strict complete-file replacement for the single target file; return pure code only; do not alter unrelated sections.\\
- Preserve interface contracts and standardized error shapes; add narrow guards instead of broad catches.\\
- Enforce cross-platform path handling and deterministic behavior; do not introduce external network calls or new side effects.\\
- Optionally add a tiny internal sanity check if it prevents recurrence without bloat.\\
- Bound attempts (<= B). If still failing, emit a compact escalation note: failing step, last command, error logs tail, and the next best single remediation.
\end{tcolorbox}

\textbf{Final Agent}
This agent consolidates artifacts and workflow logs into developer-facing documentation and delivery notes that are precise, reproducible, and integration-ready, producing a concise README with installation, quick start, key tools, troubleshooting guidance, and references.
\begin{tcolorbox}[colback=blue!5!white,colframe=blue!75!black,breakable,title=Final Agent System Prompt]
- Write a concise developer README (Markdown) including:\\
  1) Overview and value; roles of MCP and FastMCP; supported OS.\\
  2) Minimal reproducible install (Conda/venv), commands, pinned dependencies, offline notes; Windows PowerShell and Linux shell variants.\\
  3) Quick start to launch the service and call 2--3 key tools with copy-pasteable snippets and basic error handling.\\
  4) Tool list: endpoint name, one-line description, key parameters (types/defaults), return shape, idempotency/side effects, typical runtime class.\\
  5) Troubleshooting: environment/import issues, optional deps, timeouts, path problems, permissions, CPU/GPU enablement; any bounded network caveats.\\
  6) Reproducibility and telemetry: how to capture environment report, versions, minimal repro commands; where logs/artifacts live.\\
  7) License and compliance notes: repository license, usage constraints, safety guardrails.\\
- Keep structure clear, steps verifiable, and assumptions minimal; prioritize essentials for successful adoption and integration.
\end{tcolorbox}

\subsection{GPT-4 Template Baseline Prompt}
\label{sec:appendix_gpt4_template_prompt}

In the GPT-4 template baseline, we use a single system prompt that instructs one GPT-4 model to generate a complete MCP service implementation for a given repository.
The full prompt is given below.

\begin{tcolorbox}[colback=blue!5!white,colframe=blue!75!black,breakable,
                  title=GPT-4 Template Baseline System Prompt,fonttitle=\bfseries\footnotesize]
\footnotesize
You are a single GPT-4 model acting as an autonomous MCP wrapper generator for a Python code repository.

You receive:
- A short task description.
- The repository README (possibly truncated).
- A list of key source files selected by static analysis, each with its file path and content.
- A minimal description of the MCP protocol and the FastMCP library.

Your goal is to produce a single, self-contained Python MCP server that exposes a small set of high-value tools from this repository.

Requirements:
1. \textbf{Design of tools}
   - Identify 3--8 core, stable capabilities that are useful to expose as tools.
   - For each tool, choose a concise, descriptive name and a one-sentence natural-language description.
   - Specify explicit, typed parameters (no \texttt{*args} or \texttt{**kwargs}); include default values where reasonable.
   - Define clear, JSON-serializable return values. Avoid returning raw Python objects that cannot be serialized.

2. \textbf{Implementation with FastMCP}
   - Use the FastMCP library to implement the MCP server.
   - Implement a \texttt{create\_app()} function that constructs and returns the MCP application.
   - Register each tool with FastMCP, binding it to the underlying repository functions or classes.
   - Prefer importing from public, documented APIs instead of private or test modules.
   - Use lazy imports for heavy optional dependencies when necessary, and return a helpful error message if a required package is missing.

3. \textbf{Robustness and safety}
   - Validate all inputs (types, value ranges, file existence) before calling library functions.
   - Handle exceptions inside each tool and return a standardized error object instead of raising uncaught exceptions.
   - The standardized response format for each tool is:
     \texttt{\{ "success": bool, "result": any or null, "error": string or null \}}.
   - Avoid network access unless the repository explicitly requires it.
   - Restrict file I/O to the workspace; do not write to absolute paths outside the project.

4. \textbf{Coding style}
   - Write clean, idiomatic Python 3 code with clear function boundaries.
   - Add short docstrings for each tool describing its purpose, parameters, and return value.
   - Do not generate tests, examples, or documentation as part of the MCP server file.
   - Do not include Markdown fences such as `````; output only raw Python code.

5. \textbf{Output format}
   - Return a \emph{single} Python file containing the complete MCP server implementation.
   - Do not include explanations, comments to the user, or multiple files; only the final server code that can be saved as e.g. \texttt{mcp\_service.py}.
\end{tcolorbox}
\end{document}